\documentclass[review,11pt]{elsarticle}

\makeatletter
\def\ps@pprintTitle{%
 \let\@oddhead\@empty
 \let\@evenhead\@empty
 \def\@oddfoot{}%
 \let\@evenfoot\@oddfoot}
\makeatother

\makeatletter
\newcommand*\mysize{%
  \@setfontsize\mysize{9.0}{9.0}%
}
\makeatother

\makeatletter
\newcommand{\eqnum}{\refstepcounter{equation}\textup{\tagform@{\theequation}}}
\makeatother

\usepackage{comment}
\usepackage{setspace}
\usepackage[greek, english]{babel}
\usepackage{amsfonts}
\usepackage{eufrak}
\usepackage{amsmath, amsthm, amssymb}
\usepackage{palatino, url, multicol}
\usepackage{verbatim}
\usepackage{graphicx}
\usepackage{array}
\usepackage{appendix}
\usepackage{multirow}
\usepackage{subfig}
\theoremstyle{definition}

\theoremstyle{plain}

\usepackage{url}
\usepackage{tabularx}
\usepackage{ragged2e}  
\usepackage{booktabs}
\newtheorem{proposition}{Proposition}
\newtheorem{lemma}{Lemma}
\addto\captionsenglish{}
\usepackage[font=scriptsize]{caption}
\usepackage{enumitem}
\usepackage{color}
\usepackage{relsize}
\usepackage{graphicx}   	
\usepackage[left=1in,top=1in,right=1in,bottom=1in,nohead]{geometry}
\usepackage{epstopdf}
\usepackage{cases}

\begin{document}

\begin{frontmatter}

\title{Technical properties of Ranked Nodes Method}

\author[mymainaddress]{Pekka Laitila\corref{mycorrespondingauthor}}
\cortext[mycorrespondingauthor]{Corresponding author}
\ead{pekka.laitila@aalto.fi}

\author[mymainaddress,mysecondaryaddress]{Kai Virtanen}
\ead{kai.virtanen@aalto.fi}

\address[mymainaddress]{Department of Mathematics and Systems Analysis, Aalto University, Helsinki, Finland}
\address[mysecondaryaddress]{Department of Military Technology, Finnish National Defence University, Helsinki, Finland}

\date{}

\begin{abstract}
This paper presents analytical and experimental results on the ranked nodes method (RNM) that is used to construct conditional probability tables for Bayesian networks by expert elicitation. The majority of the results are focused on a setting in which RNM is applied to a child node and parent nodes that all have the same amount discrete ordinal states. The results indicate on RNM properties that can be used to support its future elaboration and development.

\end{abstract}


\end{frontmatter}


\newcommand\WMEAN{\mbox{WMEAN}}
\newcommand\WMIN{\mbox{WMIN}}
\newcommand\WMAX{\mbox{WMAX}}
\newcommand\MIXMINMAX{\mbox{MIXMINMAX}}
\section{Introduction}
Bayesian networks (BNs), see, e.g., \cite{fenton2013risk}, are directed acyclic graphs with nodes representing random variables and arcs indicating direct dependencies between them. The dependencies indicated by the arcs are quantified in conditional probability tables (CPTs). A CPT defines the probability distribution of a descendant, called a child node, for all the combinations of states of its direct predecessors, called parent nodes. Together, the CPTs and the graphical structure of a BN encode the joint probability distribution of all the nodes included. When the states of some nodes are known, the probability distributions of the others can be updated accordingly with effective algorithms, see, e.g. \cite{russell2003artificial}, \cite{neapolitan2004learning}. This feature makes BNs a useful tool to answer probabilistic queries about the nodes and thereby, provides means to support risk and decision analysis concerning them. Application areas of BNs include, e.g., medical decision making \cite{hill2021bayesian}, \cite{constantinou2015causal}, \cite{heckerman1992toward}, risk and safety management \cite{topuz2021probabilistic}, \cite{tosoni2019comprehensiveness}, as well as maintenance and policy planning \cite{mancuso2021optimal}, \cite{barons2017eliciting}.  

In practical BN applications, it is common that CPTs have to be constructed by expert elicitation due to lack of suitable data. This involves a domain expert providing subjective assessments about the probabilistic relationship of the nodes. As the size of a CPT grows exponentially with the number of parent nodes, it is easily composed of dozens or even hundreds of elements. Assessing so many probabilities directly may be impossible for the expert due to cognitive strain or lack of time \cite{druzdel2000building}, \cite{monti2000dealing}. To mitigate this challenge of expert elicitation, various techniques known as parametric methods have been developed. In these methods, the probabilistic relationship of the nodes is assumed to follow some specific rule. The rule is characterized by parameters whose number is much smaller than the number of elements in the CPT. The expert assigns values to these parameters after which the CPT can be generated for further review. A recent review of various parametric methods is found in \cite{rohmer2020uncertainties}.

This paper studies a parametric method designed for specific types of nodes called ranked nodes \cite{fenton2007using}. They represent through discrete ordinal scales quantities that can be considered continuous but for which there necessarily are no well-established interval scales. The associated parametric method is called the ranked nodes method (RNM), introduced by Fenton \textit{et al.} \cite{fenton2007using}. In RNM, three types of parameters are elicited from the expert: 1) a function called the weight expression that describes the probabilistic relationship between the parent nodes and the child node, 2) weights that represent the relative strengths of influence of the parent nodes on the child node, and 3) a variance parameter that indicates how precisely the state of the child node is known when the states of the parent nodes are known.  

RNM is implemented in AgenaRisk software \cite{agenarisk10} through which it has been applied in several real-life BN applications, see, e.g., \cite{kaya2019building}, \cite{xia2018hybrid}, \cite{freire2018bayesian}, \cite{yet2016bayesian}. In recent years, research on methodological properties of RNM has also been conducted. Application of RNM to nodes with continuous scales is elaborated in \cite{laitila2016improving} and \cite{manu4}. The theoretical principle and practical usability of the method are studied in \cite{laitila2018ontheore}. In \cite{noguchi2019addressing}, it is shown that RNM enables portraying the "explaining away" property of binary nodes more extensively than noisy-OR, a popular parametric method designed solely for binary nodes.

This paper presents on RNM analytical and experimental results that can used to support its further elaboration and development. After explaining the functioning of RNM in Section \ref{sec:CPDGen}, the paper provides in Section \ref{sec:ConvRes} an alternative form to a convegence result that characterizes the functioning of RNM. Aftert this, the paper presents a series of results specific to a setting in which a child node and all its parent nodes are ranked nodes with the same amount of states. In Section \ref{sec:Prop1Proof}, it is proven that when a conditional probability distribution is generated with RNM, the two most probable states of the child node are always consecutive on its ordinal scale. Section \ref{sec:WMEANProps} establishes results concerning a weight expression WMEAN. First, it is analytically shown that when a probability distribution of the child node is generated with WMEAN for specific types of combinations of states of the parent nodes (called scenarios), the weight of a single parent node always determines the child node's two most probable states (called the mode pair). Furthermore, through an experiment, it is concluded that critical values of this weight concerning the more pair are robust with regard to the variance parameter.  In Section \ref{sec:WMEANwu}, it is experimentally determined that probability distributions generated with WMEAN for the scenarios are robust to specific types of changes in the weights. In Section \ref{sec:WMINProp}, it is analytically shown that under certain conditions, the functional form of WMEAN coincides with that of another weight expression called WMIN. Section \ref{sec:MIXstudy} presents an experiment indicating that when a weight expression MIXMINMAX is applied in the aforementioned scenarios, a weight range implying for the child node a given mode pair is robust to the value of the variance parameter used. Concluding remarks are presented in Section \ref{sec:conclusion}.

\section{Ranked Nodes Method}\label{sec:CPDGen}

This section explains the concept of ranked nodes and the generation of CPTs with RNM. Knowledge of the generation process is necessary to understand the results on RNM established in this paper.

A ranked node is a random variable with a discrete ordinal scale such that each state can be considered to represent a range of values of a continuous quantity. The scale may consist of subjective labeled states, e.g., \{Low, Medium, High\} or it may be constructed by discretizing a continuous scale, e.g., \{[0 yr, 20 yr], [20 yr, 50 yr], [50 yr, 80 yr]\}.

\begin{figure}[h!]
\centering
\includegraphics[scale=0.25]{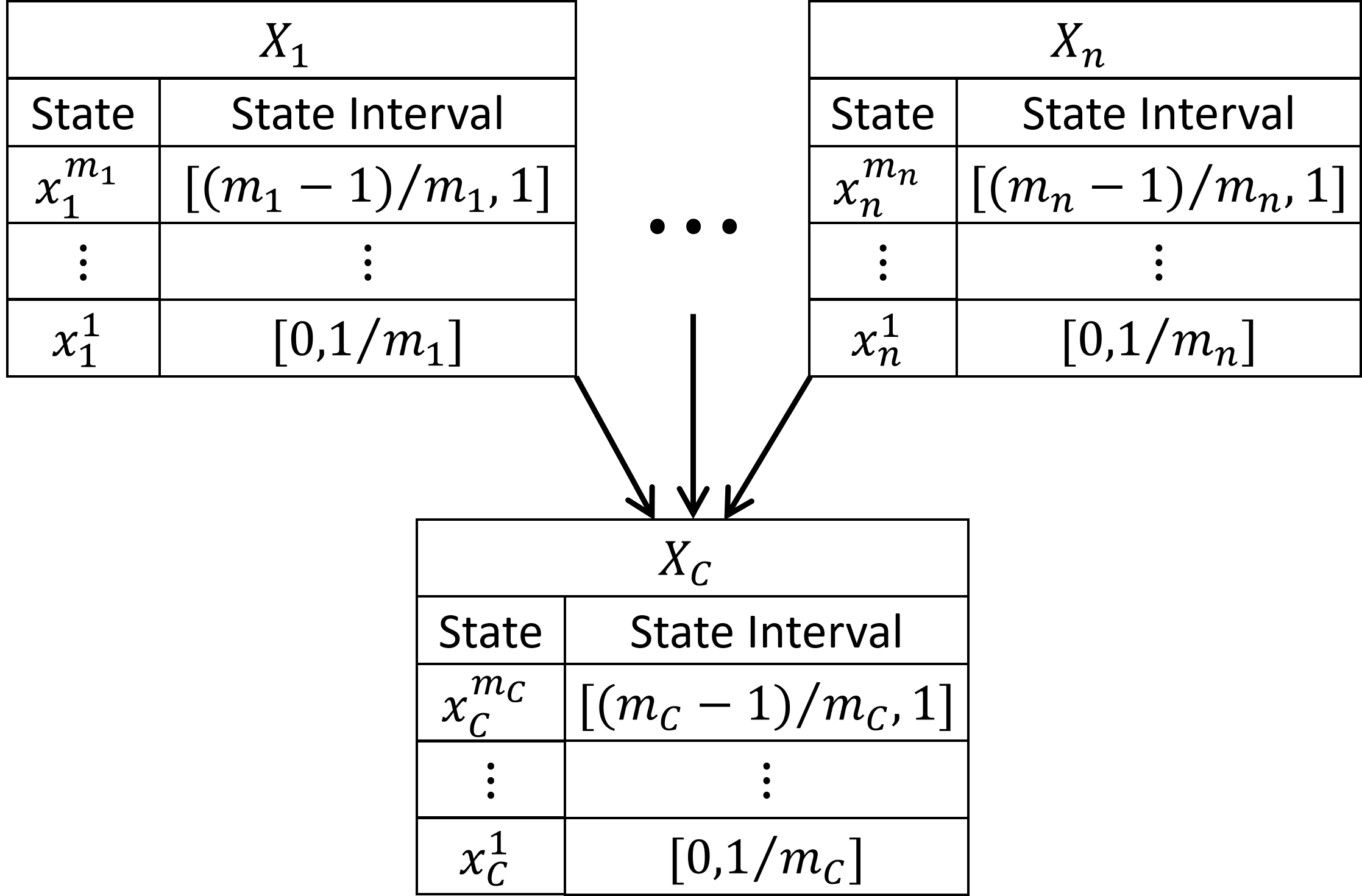}
\caption{BN fragment consisting of ranked nodes.}
\label{fig:esimverkko}
\end{figure}

To explain the functioning of RNM, consider a BN fragment consisting of parent nodes $X_1,\ldots,X_n$ and a child node $X_C$ that are all ranked nodes with $m_i$ denoting the number of states of the node $X_i$. In RNM, the consecutive states $x_i^1,\ldots,x_i^{m_i}$ of $X_i$ are associated with consecutive sub-intervals of the form $[0,1/m_i],\ldots, [(m_i-1)/m_i,1]$ on a unit scale $[0,1]$, see Fig. \ref{fig:esimverkko}. In accordance to \cite{laitila2016improving}, these sub-intervals are called state intervals from now on.

For the generation of the CPT of the child node $X_C$, the expert first selects a function $f$ known as the weight expression. It characterizes the probabilistic relationship between $X_C$ and the parent nodes $X_1,\ldots,X_n$. There are four alternative weight expressions. With WMEAN, $X_C$ represents a weighted average of $X_1,\ldots,X_n$. With WMIN, $X_C$ tends to follow parent nodes in low states but different parents can have varying strengths of influence on $X_C$. WMAX is analogical to WMIN but $X_C$ now following parent nodes in high states. In the case of MIXMINMAX, $X_C$ represents a weighted average of the lowest and highest states found among the parent nodes. After selecting the weight expressions, the expert assigns for the parent nodes weights $\boldsymbol{w}$ that reflect their relative strengths of influence on $X_C$. Finally, the expert chooses a value to a variance parameter $\sigma^2$ that indicates how precisely the state of $X_C$ is known when the states of the parent nodes are known. Once these parameters are set, the CPT can be generated. The elicitation of the parameters from the expert is discussed more in \cite{fenton2007using} and \cite{laitila2016improving}.  

With $x_i$ and $[a_i,b_i]$ denoting a given state and the associated state interval of $X_i$, the conditional probability distribution $P(X_C=x_C|X_1=x_1,\ldots,X_n=x_n)$ is generated after the parameter elicitation as follows. First, from each state interval $[a_i,b_i]$, $s$ equidistant sample points are taken so that $a_i$ and $b_i$ are among the points. Once the sample points are taken, $s^n$ unique combinations of them are formed so that in each combination, there is a sample point originating from each parent node. Let the $r$th such combination be denoted by $(z_1^r,\ldots,z_n^r)$, where $z_i^r$ is a sample point originating from the parent node $X_i$.   

With each combination of sample points $(z_1^r,\ldots,z_n^r)$, a parameter $\mu_r$ is calculated with the weight expression $f$ and the weights $\boldsymbol{w}$ by 
\begin{equation}\label{eq:muK}
\mu_r=f(z_1^r,\ldots,z_n^r, \boldsymbol{w}),
\end{equation}
where $f(z_1^r,\ldots,z_n^r, \boldsymbol{w})$ has the following functional form with different weight expressions:
\begin{align}
&\WMEAN(z_1^r,\ldots,z_n^r, \boldsymbol{w})=
\sum_{i=1}^n w_i z_i^r, \nonumber\\
&(w_1,\ldots,w_n)\in W^{\mbox{\scriptsize WMEAN}}=\{(w_1,\ldots,w_n)~|~w_1,\ldots,w_n\in[0,1],~\sum_{i=1}^n w_i=1\},\label{eq:wmean}
\end{align}
\begin{align}
&\WMIN(z_1^r,\ldots,z_n^r, \boldsymbol{w})=\min\limits_{i=1,\ldots,n}\left\{\frac{w_i z_i^r+\sum_{j\neq i}z_j^r}{w_i+n-1}\right\}, \nonumber\\
&(w_1,\ldots,w_n)\in W^{\mbox{\scriptsize WMIN}}=\{(w_1,\ldots,w_n)~|~w_1,\ldots,w_n \geq 1\},\label{eq:wmin}
\\
\addlinespace
\addlinespace
\addlinespace
&\WMAX(z_1^r,\ldots,z_n^r, \boldsymbol{w})=
\max\limits_{i=1,\ldots,n}\left\{\frac{w_i z_i^r+\sum_{j\neq i}z_j^r}{w_i+n-1}\right\},\nonumber\\
&(w_1,\ldots,w_n)\in W^{\mbox{\scriptsize WMAX}}=\{(w_1,\ldots,w_n)~|~w_1,\ldots,w_n \geq 1\}, \label{eq:wmax} 
\\
\addlinespace
\addlinespace
\addlinespace
&\MIXMINMAX(z_1^r,\ldots,z_n^r, \boldsymbol{w})=
w_{MIN}*\min\limits_{i=1,\ldots,n}\{z_i^r\}+w_{MAX}*\max\limits_{i=1,\ldots,n}\{z_i^r\}, \nonumber\\
&(w_{MIN},w_{MAX})\in W^{\mbox{\scriptsize MIXMINMAX}}=\{(w_{MIN},w_{MAX})~|~w_{MIN}\in[0,1],\nonumber \\
& \quad w_{MAX}=1-w_{MIN}\}.\label{eq:mixminmax} 
\end{align}
In Eqs. \eqref{eq:wmean}  -- \eqref{eq:mixminmax}, $W^f$ is the set of feasible weights regarding the weight expression $f$. 

With each $\mu_r$, a value $p_{[a_C,b_C]}^{\mu_r}$ is calculated with
\begin{align}\label{apu0}
p_{[a_C,b_C]}^{\mu_r}=\int_{a_C}^{b_C} TNormpdf(u,\mu_r,\sigma^2,0,1)du,
\end{align} 
where $\sigma^2$ is the variance parameter and $TNormpdf(u,\mu_r,\sigma^2,0,1)$ is the probability density function of a normal distribution $N(\mu_r,\sigma^2)$ truncated to $[0,1]$, i.e.,
\begin{align}
TNormpdf(u,\mu_r,\sigma^2,0,1)=\frac{\frac{1}{\sqrt{2\pi}\sigma}e^{-\frac{(u-\mu_r)^2}{2\sigma^2}}}{\int_{0}^{1}\frac{1}{\sqrt{2\pi}\sigma}e^{-\frac{(u-\mu_r)^2}{2\sigma^2}}du}.
\end{align} 
After this, the probability $P(X_C=x_C|X_1=x_1,\ldots,X_n=x_n)$ is formed as the average
\begin{equation}\label{eq:PcX}
P(X_C=x_C|X_1=x_1,\ldots,X_n=x_n)=\frac{\sum_{r=1}^{s^n}p_{[a_C,b_C]}^{\mu_r}}{s^n}.
\end{equation}

The whole CPT of the child node is obtained by repeating
the computations presented above for all states of the
child node with each combination of the states of the parent
nodes.

It is shown in \cite{laitila2018ontheore} that as the sample size $s$ increases, the conditional probabilities generated with RNM approximate with an increasing accuracy conditional probabilities related to a regression model of continuous random variables. The result is presented in \cite{laitila2018ontheore} as follows. Let $P(X_C=x_C|X_1=x_1,\ldots,X_n=x_n; f,\boldsymbol{w},\sigma^2,s)$ denote a conditional probability generated with RNM using the weight expression $f$, the weights $\boldsymbol{w}$, the variance parameter $\sigma^2$, and the sample size $s$. Then,
\begin{align}\label{eq:RNMprincip}
&P(X_C=x_C|X_1=x_1,\ldots,X_n=x_n; f,\boldsymbol{w},\sigma^2,s)\xrightarrow{s\rightarrow \infty} \nonumber\\
&P(\chi_C\in[a_C,b_C]~|~\chi_1\sim U(a_1,b_1),\ldots,\chi_n\sim U(a_n,b_n),\chi_C\in[0,1]),
\end{align}  
where  $\chi_1,\ldots\chi_n,\chi_C$ are continuous random variables and $\chi_C$ depends on $\chi_1,\ldots\chi_n$ according to the regression model
\begin{equation}\label{eq:regModel}
\chi_C=f(\chi_1,\ldots,\chi_n,\boldsymbol{w})+e,~e\sim N(0,\sigma^2).
\end{equation}

\section{Alternative form for converge result concerning RNM}\label{sec:ConvRes}
An alternative form of the convergence result \eqref{eq:RNMprincip} is next derived. It is referred to in the paper later on. Let the random variable $\mu$ be defined as
\begin{align}\label{eq:mu}
\mu=f(\chi_1,\ldots,\chi_n,\boldsymbol{w}),~\chi_i\sim U(a_i,b_i)~\forall~i=1,\ldots,n.
\end{align}
Then, $\chi_C$ defined in Eq. \eqref{eq:regModel} can be written as $\chi_C=\mu+e$. When $\mu$ is fixed to a given value $\mu=u$, the probability distribution of $\chi_C$ follows a normal distribution with $u$ as the mean and $\sigma^2$ as the variance. That is, $\chi_C|(\mu=u)\sim N(u,\sigma^2)$. If it is also known that $\chi_C$ lies within $[0,1]$, the probability distribution of $\chi_C$ corresponds $N(u,\sigma^2)$ truncated to $[0,1]$. In such a case, it can therefore be written
\begin{equation}
P(\chi_C\in[a_C,b_C]~|~\mu=u,\chi_C\in[0,1])=\int_{a_C}^{b_C}TNormpdf(x,u,\sigma^2,0,1)dx.
\end{equation} 
Now, by utilising the definition of $\mu$ in Eq. \eqref{eq:mu} and the above equation, it can be written
\begin{align}\label{eq:RNMprincipApu}
&P(\chi_C\in[a_C,b_C]~|~\chi_1\sim U(a_1,b_1),\ldots,\chi_n\sim U(a_n,b_n),\chi_C\in[0,1])\nonumber\\
&=\int_{-\infty}^{\infty}P(\chi_C\in[a_C,b_C]~|~\mu=u,\chi_C\in[0,1])f_{\mu}(u)du\nonumber\\
\addlinespace
&=\int_{-\infty}^{\infty}\left[\int_{a_C}^{b_C}TNormpdf(x,u,\sigma^2,0,1)dx\right]f_{\mu}(u)du\nonumber\\
\addlinespace
&=\int_{\underline{\mu}}^{\overline{\mu}}\left[\int_{a_C}^{b_C}TNormpdf(x,u,\sigma^2,0,1)dx\right] f_{\mu}(u)du,
\end{align}
where $f_{\mu}$ denotes the probability density function of $\mu$ while $\underline{\mu}$ and $\overline{\mu}$ correspond to the lower and upper bounds of $\mu$, i.e.,
\begin{equation}\label{eq:muBounds}
\underline{\mu}=\min_{\substack{\chi_i\in[a_i,b_i]\\\forall i=1,\ldots,n}}\{f(\chi_1,\ldots,\chi_n,\boldsymbol{w})\}, \qquad \overline{\mu}=\max_{\substack{\chi_i\in[a_i,b_i]\\\forall i=1,\ldots,n}}\{f(\chi_1,\ldots,\chi_n,\boldsymbol{w})\}.
\end{equation}

Based on Eq. \eqref{eq:RNMprincipApu}, the convergence result \eqref{eq:RNMprincip} can be presented as 
\begin{align}\label{eq:RNMprincip2}
&P(X_C=x_C|X_1=x_1,\ldots,X_n=x_n; f,\boldsymbol{w},\sigma^2,s)\xrightarrow{s\rightarrow \infty} \nonumber\\
&\int_{\underline{\mu}}^{\overline{\mu}}\left[\int_{a_C}^{b_C}TNormpdf(x,u,\sigma^2,0,1)dx\right] f_{\mu}(u)du.
\end{align} 

\section{Two most probable states consecutive in probability distributions generated with RNM}\label{sec:Prop1Proof}
In this section, the following feature of conditional probability distributions generated with RNM is proved.
\begin{proposition}\label{prop1}
Let there be parent nodes $X_1,\ldots,X_n$ and a child node $X_C$ each with $m$ states. Let a CPT be generated for $X_C$ by using RNM with a given weight expression $f\in\{\WMEAN, \WMIN, \WMAX,\linebreak \MIXMINMAX \}$ and given values of the weights $\boldsymbol{w}\in W^f$, the variance parameter $\sigma^2>0$, and the sample size $s\geq 3$. Then, for any conditional probability distribution $P(X_C|X_1=x_1,\ldots,X_n=x_n)$, it applies that the two most probable states of $X_C$ are consecutive on its ordinal scale.
\end{proposition}

The proof of Proposition \ref{prop1} is presented below after two aiding lemmas. 

\begin{lemma}\label{lemma1}
Let there be parent nodes $X_1,\ldots,X_n$ and a child node $X_C$ each with $m$ states. Let a conditional probability distribution $P(X_C|X_1=x_1,\ldots,X_n=x_n)$ be generated with RNM using a given weight expression $f\in\{\WMEAN, \WMIN, \WMAX, \MIXMINMAX \}$ and given values of the weights $\boldsymbol{w}\in W^f$, the variance parameter $\sigma^2>0$, and the sample size $s\geq 3$. Then, it applies
\begin{equation}\label{eq:mu_range}
\max_{r=1,\ldots,s^n}\{\mu_r\}-\min_{r=1,\ldots,s^n}\{\mu_r\}=\frac{1}{m},
\end{equation}
where $\mu_r$, $r=1,\ldots,s^n$ are parameters calculated in the generation process according to Eq. \eqref{eq:muK}.
\end{lemma}
\begin{proof}
Condition \eqref{eq:mu_range} is shown to be valid separately for each weight expression. Without the loss of generality, assume that the given state $x_i$ of the parent node $X_i$ is its $k_i$th state on the ordinal scale. The state interval associated with this state is then $[\frac{k_i-1}{m},\frac{k_i}{m}]$. 

\textbf{WMEAN}

By taking a partial derivative of Eq. \eqref{eq:wmean}, it is obtained
\begin{equation}
\frac{\partial WMEAN}{\partial z_i^r}=w_i\geq 0
\end{equation}
which indicates that WMEAN is an increasing function with respect to any sample point $z_i^r$. Thus, $\min\{\mu_r\}_{r=1}^{s^n}$ is obtained with a sample point combination $(z_1^r,\ldots,z_n^r)$ in which each individual sample point $z_i^r$ is in its smallest possible value, i.e., $z_i^r=(k_i-1)/m$. Correspondingly, $\max\{\mu_r\}_{r=1}^{s^n}$ is obtained with a combination in which each individual sample point is in its largest possible value, i.e., $z_i^r=k_i/m$. Therefore, one obtains
\begin{align}
\max_{r=1,\ldots,s^n}\{\mu_r\}-\min_{r=1,\ldots,s^n}\{\mu_r\}&=\sum_{i=1}^n w_i*\max_{j=1,\ldots,s}\{z_i^j\}-\sum_{i=1}^n w_i*\min_{j=1,\ldots,s}\{z_i^j\}\nonumber\\
&=\sum_{i=1}^n w_i*\left(\frac{k_i}{m}-\frac{k_i-1}{m}\right)\nonumber \\
&=\frac{1}{m}\sum_{i=1}^n w_i\nonumber\\
&=\frac{1}{m}.
\end{align}
 
\textbf{WMIN}

Referring to Eqs. \eqref{eq:muK} and \eqref{eq:wmin}, let $\mu_r^i$ be defined as
\begin{equation}
\mu_r^i=\frac{w_i z_i^r+\sum_{j\neq i}z_j^r}{w_i+n-1},
\end{equation} 
which enables one to represent $\mu_r$ given by WMIN as
\begin{equation}
\mu_r=WMIN(z_1^r,\ldots,z_n^r,w_1,\ldots,w_n)=\min_{i=1,\ldots,n}\{\mu_r^i\}.
\end{equation}
Now
\begin{equation}
\frac{\partial \mu_r^i}{\partial z_i^r}=\frac{w_i}{w_i+n-1}>0,
\end{equation}
which indicates that $\mu_r^i$  is a strictly increasing function with respect to $z_i^r$. Thus, like with WMEAN, $\min\{\mu_r\}_{r=1}^{s^n}$ is obtained with a sample point combination $(z_1^r,\ldots,z_n^r)$ in which each individual sample point $z_i^r$ is in its smallest possible value. Correspondingly, $\max\{\mu_r\}_{r=1}^{s^n}$ is obtained with a combination in which each individual sample point is in its largest possible value. Thus, it is obtained
\begin{align}
&\max_{r=1,\ldots,s^n}\{\mu_r\}-\min_{r=1,\ldots,s^n}\{\mu_r\}\nonumber\\
&=\min_{i=1,\ldots,n}\left\{\frac{w_i*\max\limits_{j=1,\ldots,s}\{z_i^j\}+\sum_{t\neq i}\max\limits_{j=1,\ldots,s}\{z_t^j\}}{w_i+n-1}\right\}\nonumber\\
&-\min_{i=1,\ldots,n}\left\{\frac{w_i*\min\limits_{j=1,\ldots,s}\{z_i^j\}+\sum_{t\neq i}\min\limits_{j=1,\ldots,s}\{z_t^j\}}{w_i+n-1}\right\}\nonumber
\end{align}
\begin{align}
&=\min_{i=1,\ldots,n}\left\{\frac{w_i*\frac{k_i}{m}+\sum_{t\neq i}\frac{k_t}{m}}{w_i+n-1}\right\}-\min_{i=1,\ldots,n}\left\{\frac{w_i*\frac{k_i-1}{m}+\sum_{t\neq i}\frac{k_t-1}{m}}{w_i+n-1}\right\}\nonumber\\
&=\frac{1}{m}.
\end{align}

\textbf{WMAX}

With WMAX, Eq. \eqref{eq:mu_range} can be shown to be true by following the same steps as with WMIN above. 

\textbf{MIXMINMAX}

With MIXMINMAX, it can be derived straight from Eqs. \eqref{eq:muK} and \eqref{eq:mixminmax} that
\begin{align}
\max_{r=1,\ldots,s^n}\{\mu_r\}-\min_{r=1,\ldots,s^n}\{\mu_r\}&=
\max_{r=1,\ldots,s^n}\{w_{MIN}*\min\limits_{i=1,\ldots,n}\{z_i^r\}+w_{MAX}*\max\limits_{i=1,\ldots,n}\{z_i^r\}\}\nonumber\\
&-\min_{r=1,\ldots,s^n}\{w_{MIN}*\min\limits_{i=1,\ldots,n}\{z_i^r\}+w_{MAX}*\max\limits_{i=1,\ldots,n}\{z_i^r\}\}\nonumber\\
&=w_{MIN}*\frac{\min\limits_{i=1,\ldots,n}\{k_i\}}{m}+w_{MAX}*\frac{\max\limits_{i=1,\ldots,n}\{k_i\}}{m}\nonumber\\
&-w_{MIN}*\frac{\min\limits_{i=1,\ldots,n}\{k_i\}-1}{m}-w_{MAX}*\frac{\max\limits_{i=1,\ldots,n}\{k_i\}-1}{m}\nonumber\\
&=w_{MIN}*\frac{1}{m}+w_{MAX}*\frac{1}{m}\nonumber\\
&=\frac{1}{m}.
\end{align}
\end{proof}  
\pagebreak
\begin{lemma}\label{lemma2}
Let $[0, 1]$ be divided into $m$ subintervals of the form $[\frac{i-1}{m},\frac{i}{m}],~ i=1,...,m$. Furthermore, let $\mu\in[0,1]$. Then, with any $i,j=1,\ldots,m,~j\neq i$, it applies
\begin{align}\label{eq:lemma2Claim}
&|\tfrac{i-1/2}{m}-\mu|\leq |\tfrac{j-1/2}{m}-\mu| \Rightarrow \nonumber\\ 
&\int_{\frac{i-1}{m}}^{\frac{i}{m}}TNormpdf(u,\mu,\sigma^2,0,1)du \geq \int_{\frac{j-1}{m}}^{\frac{j}{m}}TNormpdf(u,\mu,\sigma^2,0,1)du.
\end{align}
\end{lemma}
\begin{proof}

Suppose first that $\mu\notin[\frac{i-1}{m},\frac{i}{m}]$ and $\mu\notin[\frac{j-1}{m},\frac{j}{m}]$. Let both of the intervals be divided into $t$ parts of equal width. For both cases $k=i$ and $k=j$, let the division be characterized by points of the form $c_k+d_k^r$, $r=1,\ldots,t$ where $c_k$ and $d_k^r$ are defined as
\begin{equation}
\left\{ \begin{array}{ll}
 c_k=\frac{k-1}{m},~d_k^r=\frac{r}{mt} &\mbox{ if $\mu < \frac{k-1}{m}$} \\
  c_k=\frac{k}{m},~d_k^r=-\frac{r}{mt} &\mbox{ if $\mu > \frac{k}{m}$}
       \end{array} \right..
\end{equation}
By the definitions of $c_k$ and $d_k^r$, it applies in both cases $k=i$ and $k=j$ that
\begin{align}
|c_k+d_k^r-\mu|=|c_k-\mu|+\frac{r}{mt}=|\tfrac{k-1/2}{m}-\mu|-\frac{1/2}{m}+\frac{r}{mt}, 
\end{align}
from which it follows
\begin{equation}\label{eq:apu11}
|\tfrac{i-1/2}{m}-\mu|\leq |\tfrac{j-1/2}{m}-\mu| \Rightarrow |c_i+d_i^r-\mu|\leq|c_j+d_j^r-\mu|.
\end{equation}
It can readily be verified that
\begin{equation}\label{eq:apu22}
|u_1-\mu|\leq |u_2-\mu|\Rightarrow TNormpdf(u_1,\mu,\sigma^2,0,1) \geq TNormpdf(u_2,\mu,\sigma^2,0,1).
\end{equation}
Combining Eqs. \eqref{eq:apu11} and \eqref{eq:apu22}, and using the fact that $TNormpdf(u,\mu,\sigma^2,0,1)$ is an integrable function \cite{knapp2005basic}, it is then obtained
\begin{align}\label{eq:apu33}
&\int_{\frac{i-1}{m}}^{\frac{i}{m}}TNormpdf(u,\mu,\sigma^2,0,1)du - \int_{\frac{j-1}{m}}^{\frac{j}{m}}TNormpdf(u,\mu,\sigma^2,0,1)du \nonumber\\
&=\lim_{t\rightarrow \infty}\sum_{r=1}^t TNormpdf(c_i+d_i^r,\mu,\sigma^2,0,1)*\frac{1}{mt}\nonumber\\
&-\lim_{t\rightarrow \infty}\sum_{r=1}^t TNormpdf(c_j+d_j^r,\mu,\sigma^2,0,1)*\frac{1}{mt}\nonumber\\
&=\lim_{t\rightarrow \infty}\sum_{r=1}^t\underbrace{\left(TNormpdf(c_i+d_i^r,\mu,\sigma^2,0,1)-TNormpdf(c_j+d_j^r,\mu,\sigma^2,0,1) \right)}_{\geq 0}*\frac{1}{mt}\nonumber\\
&\geq 0,
\end{align}
by which Eq. \eqref{eq:lemma2Claim} is proved under the assumptions $\mu\notin[\frac{i-1}{m},\frac{i}{m}]$ and $\mu\notin[\frac{j-1}{m},\frac{j}{m}]$.

Next, suppose that $\mu\in[\frac{i-1}{m},\frac{i}{m}]$. Then, 
it necessarily holds that
\begin{equation}
|\tfrac{i-1/2}{m}-\mu|\leq \frac{1/2}{m},\qquad |\tfrac{j-1/2}{m}-\mu|\geq \frac{1/2}{m}~\forall~j=1,\ldots,m,~j\neq i.
\end{equation}
That is, the condition $\mu\in[\frac{i-1}{m},\frac{i}{m}]$ is a special case of the condition $|\tfrac{i-1/2}{m}-\mu|\leq |\tfrac{j-1/2}{m}-\mu|$. Let $c_i$, $d_i^r$, $c_j$, and $d_j^r$ now be defined by
\begin{equation}
i > j \Rightarrow \left\{ \begin{array}{ll}
 c_i=\frac{i-1}{m},~d_i^r=\frac{r}{mt} \\
  c_j=\frac{j}{m},~d_j^r=-\frac{r}{mt}
       \end{array} \right.,\\
     i < j \Rightarrow \left\{ \begin{array}{ll}
 c_i=\frac{i}{m},~d_i^r=-\frac{r}{mt} \\
  c_j=\frac{j-1}{m},~d_j^r=\frac{r}{mt}
       \end{array} \right..  
\end{equation}
Then, it necessarily follows that 
\begin{equation}
|c_i+d_i^r-\mu|\leq|c_j+d_j^r-\mu|,
\end{equation}
which together with Eqs. \eqref{eq:apu22} and \eqref{eq:apu33} implies
\begin{equation}
\int_{\frac{i-1}{m}}^{\frac{i}{m}}TNormpdf(u,\mu,\sigma^2,0,1)du - \int_{\frac{j-1}{m}}^{\frac{j}{m}}TNormpdf(u,\mu,\sigma^2,0,1)du\geq 0.
\end{equation}
Thus, Eq. \eqref{eq:lemma2Claim} is now proved also in the special case $\mu\in[\frac{i-1}{m},\frac{i}{m}]$ of the given condition $|\tfrac{i-1/2}{m}-\mu|\leq |\tfrac{j-1/2}{m}-\mu|$ and thereby, the whole proof is complete. 
\end{proof}

\textbf{Proof of Proposition \ref{prop1}}

In the proof of Lemma \ref{lemma1}, WMEAN, WMIN, and WMAX are shown to be increasing functions with respect to the sample points $z_i^r$ taken from the state intervals of the parent nodes, see also Eqs. \eqref{eq:muK} -- \eqref{eq:wmax}. This property is also shared by MIXMINMAX as it is a weighted sum of the increasing $\min$ and $\max$ functions, see Eq. \eqref{eq:mixminmax}. Therefore, regarding $\mu$ defined in Eq. \eqref{eq:mu}, the bounds $\underline{\mu}$ and $\overline{\mu}$ defined in Eq. \eqref{eq:muBounds} can also equivalently be written as 
\begin{equation}\label{eq:muBounds2}
\underline{\mu}=\min_{r=1,\ldots,s^n}\{\mu_r\}, \qquad \overline{\mu}=\max_{r=1,\ldots,s^n}\{\mu_r\}.
\end{equation}
By Lemma \ref{lemma1}, it is known that 
\begin{equation}
\overline{\mu}-\underline{\mu}=\frac{1}{m}.
\end{equation}

With $i=1,\ldots,m$, let $x_C^i$ denote the $i$th state of the child node and let $z_i=[\frac{i-1}{m},\frac{i}{m}]$ denote the corresponding state interval. Furthermore, referring to Eq. \eqref{apu0}, for $r=1,\ldots,s^n$ and $i=1,\ldots,m$, let the value $p_{z_i}^{\mu_r}$ be defined as
\begin{equation}\label{eq:pApu}
p_{z_i}^{\mu_r}=\int_{\frac{i-1}{m}}^{\frac{i}{m}}TNormpdf(u,\mu_r,\sigma^2,0,1)du.
\end{equation}
Now, suppose first that with some $k\in \mathbb{N},~1\leq k\leq m$
\begin{equation}\label{eq:muCond1}
\underline{\mu}\in(\tfrac{k-1}{m},\tfrac{k-1/2}{m}].
\end{equation}
Let the set $\Omega=\{\mu_r\}_{r=1}^{s^n}$ be divided into three subsets $\Omega_1$, $\Omega_2$, and $\Omega_3$ defined by
\begin{align}\label{eq:omegaDef}
&\Omega_1=\{\mu_r\in\Omega~|~\mu_r\in[\underline{\mu},
\tfrac{k-1/2}{m}]\},\nonumber\\
\addlinespace
&\Omega_2=\{\mu_r\in\Omega~|~\mu_r\in(\tfrac{k-1/2}{m},\tfrac{k}{m}]\},\nonumber\\
\addlinespace
&\Omega_3=\{\mu_r\in\Omega~|~\mu_r\in(\tfrac{k}{m},\overline{\mu}]\}.
\end{align}
Based on Lemma \ref{lemma2} and Eq. \eqref{eq:pApu}, it applies
\begin{align}
&\mu_r \in\Omega_1\Rightarrow p_{z_k}^{\mu_r}\geq p_{z_{k-1}}^{\mu_r} \geq p_{z_{k+1}}^{\mu_r} \geq p_{z_{i}}^{\mu_r}~~\forall~i=1,\ldots,m,i\neq k-1,k,k+1, \nonumber\\
\addlinespace
&\mu_r \in\Omega_2\Rightarrow p_{z_k}^{\mu_r}\geq p_{z_{k+1}}^{\mu_r} \geq p_{z_{k-1}}^{\mu_r} \geq p_{z_{i}}^{\mu_r}~~\forall~i=1,\ldots,m,i\neq k-1,k,k+1,
\nonumber\\
\addlinespace
&\mu_r \in\Omega_3\Rightarrow p_{z_{k+1}}^{\mu_r}\geq p_{z_{k}}^{\mu_r} \geq p_{z_{k+2}}^{\mu_r} \geq p_{z_{i}}^{\mu_r}~~\forall~i=1,\ldots,m,i\neq k,k+1,k+2.
\end{align}
As a result, for probabilities generated with RNM, see Eq. \eqref{eq:PcX}, it is obtained
\begin{align}\label{eq:Pero1}
&P(X_C=x_C^k|X_1=x_1,\ldots,X_n=x_n)-P(X_C=x_C^i|X_1=x_1,\ldots,X_n=x_n)\nonumber\\
\addlinespace
&=\frac{\sum\limits_{\mu_r \in\Omega_1}p_{z_k}^{\mu_r}+\sum\limits_{\mu_r \in\Omega_2}p_{z_k}^{\mu_r}+\sum\limits_{\mu_r \in\Omega_3}p_{z_k}^{\mu_r}}{s^n}-\frac{\sum\limits_{\mu_r \in\Omega_1}p_{z_i}^{\mu_r}+\sum\limits_{\mu_r \in\Omega_2}p_{z_i}^{\mu_r}+\sum\limits_{\mu_r \in\Omega_3}p_{z_i}^{\mu_r}}{s^n}\nonumber\\
\addlinespace
&=\frac{\sum\limits_{\mu_r \in\Omega_1}\underbrace{\left(p_{z_k}^{\mu_r}-p_{z_i}^{\mu_r} \right)}_{\geq 0}+\sum\limits_{\mu_r \in\Omega_2}\underbrace{\left(p_{z_k}^{\mu_r}-p_{z_i}^{\mu_r} \right)}_{\geq 0}+\sum\limits_{\mu_r \in\Omega_3}\underbrace{\left(p_{z_k}^{\mu_r}-p_{z_i}^{\mu_r} \right)}_{\geq 0}}{s^n}\nonumber\\
&\geq 0~\forall~i=1,\ldots,m,~i\neq k,k+1.
\end{align}
Thus, it is concluded that $x_C^k$ must be among the two most probable states of $X_C$ in the conditional probability distribution $P(X_C|X_1=x_1,\ldots,X_n=x_n)$. In a similar manner it can be concluded that 
\begin{align}\label{eq:Pero2}
&P(X_C=x_C^{k+1}|X_1=x_1,\ldots,X_n=x_n)-P(X_C=x_C^i|X_1=x_1,\ldots,X_n=x_n)\geq 0\nonumber\\
\addlinespace
&\quad\forall~i=1,\ldots,m,~i\neq k,k-1.
\end{align}
Together Eqs. \eqref{eq:Pero1} and \eqref{eq:Pero2} imply that when  when $\underline{\mu}\in(\tfrac{k-1}{m},\tfrac{k-1/2}{m}]$, the two most probable states of $X_C$ ordered with the mode first is always one of the pairs $(x_C^k,x_C^{k+1})$, $(x_C^{k+1},x_C^{k})$, or $(x_C^k,x_C^{k-1})$. 

Consider then the case
\begin{equation}\label{muCond2}
\underline{\mu}=\tfrac{k-1}{m}.
\end{equation}
In this specific case, the evaluations in Eqs. \eqref{eq:omegaDef} -- \eqref{eq:Pero1} can be repeated so that $\Omega_1$ and $\Omega_2$ are defined as above whereas $\Omega_3$ is defined as an empty set. Then, the result is that the two most probable states of $X_C$ ordered with the mode first is necessarily either of the pairs $(x_C^k,x_C^{k+1})$ and $(x_C^k,x_C^{k-1})$. 

Suppose then that 
\begin{equation}
\underline{\mu}\in(\tfrac{k-1/2}{m},\tfrac{k}{m}].
\end{equation} 
Let the set $\Omega=\{\mu_r\}_{r=1}^{s^n}$ be divided into three subsets $\Omega_1$, $\Omega_2$, and $\Omega_3$ defined now by
\begin{align}
&\Omega_1=\{\mu_r\in\Omega~|~\mu_r\in[\underline{\mu},
\tfrac{k}{m}]\},\nonumber\\
\addlinespace
&\Omega_2=\{\mu_r\in\Omega~|~\mu_r\in(\tfrac{k}{m},\tfrac{k+1/2}{m}]\},\nonumber\\
\addlinespace
&\Omega_3=\{\mu_r\in\Omega~|~\mu_r\in(\tfrac{k+1/2}{m},\overline{\mu}]\}.
\end{align}

Based on Lemma \ref{lemma2}, it then applies
\begin{align}
&\mu_r \in\Omega_1\Rightarrow p_{z_k}^{\mu_r}\geq p_{z_{k+1}}^{\mu_r} \geq p_{z_{k-1}}^{\mu_r} \geq p_{z_{i}}^{\mu_r}~~\forall~i=1,\ldots,m,i\neq k-1,k,k+1, \nonumber\\
\addlinespace
&\mu_r \in\Omega_2\Rightarrow p_{z_{k+1}}^{\mu_r}\geq p_{z_{k}}^{\mu_r} \geq p_{z_{k+2}}^{\mu_r} \geq p_{z_{i}}^{\mu_r}~~\forall~i=1,\ldots,m,i\neq k,k+1,k+2,
\nonumber\\
\addlinespace
&\mu_r \in\Omega_3\Rightarrow p_{z_{k+1}}^{\mu_r}\geq p_{z_{k+2}}^{\mu_r} \geq p_{z_{k}}^{\mu_r} \geq p_{z_{i}}^{\mu_r}~~\forall~i=1,\ldots,m,i\neq k,k+1,k+2.
\end{align}
Thereby, through similar calculations as in Eq. \eqref{eq:Pero1}, it is obtained that  
\begin{align}\label{eq:Pero3}
&P(X_C=x_C^{k+1}|X_1=x_1,\ldots,X_n=x_n)-P(X_C=x_C^i|X_1=x_1,\ldots,X_n=x_n)\geq 0\nonumber\\
\addlinespace
&\quad\forall~i=1,\ldots,m,~i\neq k,k+1, \nonumber
\end{align}
\begin{align}
&P(X_C=x_C^{k}|X_1=x_1,\ldots,X_n=x_n)-P(X_C=x_C^i|X_1=x_1,\ldots,X_n=x_n)\geq 0\nonumber\\
\addlinespace
&\quad\forall~i=1,\ldots,m,~i\neq k,k+1,k+2,
\end{align}
which implies that when $\underline{\mu}\in(\tfrac{k-1/2}{m},\tfrac{k}{m}]$, the two most probable states of $X_C$ ordered with the mode first is necessarily one of the pairs $(x_C^{k},x_C^{k+1})$, $(x_C^{k+1},x_C^{k})$, or $(x_C^{k+1},x_C^{k+2})$.

Above, the conditions $\underline{\mu}\in(\tfrac{k-1}{m},\tfrac{k-1/2}{m}]$, $\underline{\mu}=\tfrac{k-1}{m}$, and $\underline{\mu}\in(\tfrac{k-1/2}{m},\tfrac{k}{m}]$ are separately addressed. As a whole it is shown that when $\underline{\mu}\in[\tfrac{k-1}{m},\tfrac{k}{m}]$ with some $k=1,\ldots,m$, the most and the second most probable states of the child node correspond to one of the following pairs of consecutive states of the child node: $(x_C^{k},x_C^{k+1})$, $(x_C^{k+1},x_C^{k})$, $(x_C^{k},x_C^{k-1})$, or $(x_C^{k+1},x_C^{k+2})$. As the condition $\underline{\mu}\in[\tfrac{k-1}{m},\tfrac{k}{m}]$ necessarily holds with some specific $k=1,\ldots,m$, Proposition \ref{prop1} is proved.

\section{Connection between weights and mode pairs obtained with WMEAN}\label{sec:WMEANProps}

This section studies properties of probability distributions generated with WMEAN. It is analytically shown that in specific settings, the weight of a single parent node determines the mode pair, i.e, the two most probable states of the child node, independent of the weights of the other parent nodes. Furthermore, it is experimentally determined that critical values of this single weight concerning the mode pair are robust with regard to the variance parameter of RNM. 

Let there be parent nodes $X_1,\ldots,X_n$ and a child node $X_C$ that all have $m$ states. Consider a scenario in which the parent node $X_i$ is in its lowest state $x_i^1$ and the rest of the parent nodes $X_j$, $j\neq i$, are in their highest states $x_j^m$. In what follows, this scenario is denoted by $x^{D,i}$ defined by
\begin{equation}\label{eq:sceDi}
x^{D,i}=(x^{D,i}_1,\ldots,x^{D,i}_n), \qquad\left\{\begin{array}{lr}
x^{D,i}_j=x_j^{m}~~\forall~j=1,\ldots,n,~j\neq i \\
x^{D,i}_i=x_i^{1}
\end{array}
\right..
\end{equation} 
Let the conditional probability distribution $P(X_C|x^{D,i})$ be generated with RNM using WMEAN with the weights $(w_1,\ldots,w_n)\in W^{\mbox{\scriptsize WMEAN}}$ and the variance parameter $\sigma^2$. Then, for the probability of a given state $x_C^k$, it applies, see Eq. \eqref{eq:RNMprincip2},
\begin{align}\label{eq:WMEANDscen}
&P(X_C=x_C^k|x^{D,i}; \WMEAN,(w_1,\ldots,w_n),\sigma^2,s)\xrightarrow{s\rightarrow \infty} \nonumber\\
&\int_{\underline{\mu}}^{\overline{\mu}}\left[\int_{\frac{k-1}{m}}^{\frac{k}{m}}TNormpdf(x,u,\sigma^2,0,1)dx\right] f_{\mu}(u)du,
\end{align}
where $\mu$ defined in Eq. \eqref{eq:mu} has now the form
\begin{align}\label{eq:muDwmean}
&\mu=\sum_{j=1}^n w_j \chi_j,~\chi_j\sim U(\tfrac{m-1}{m},1)~\forall~j=1,\ldots,n,~j\neq i,~\chi_i\sim U(0,\tfrac{1}{m}),\nonumber\\
&w_j\in[0,1]~\forall~j=1,\ldots,n,~~\sum_{j=1}^n w_j=1,
\end{align}
and the values of $\underline{\mu}$ and $\overline{\mu}$ defined in Eqs. \eqref{eq:muBounds} and \eqref{eq:muBounds2} are now 
\begin{align}\label{eq:muBounds3}
&\underline{\mu}=\sum_{j\neq i}^n w_j \frac{m-1}{m}=\frac{(1-w_i)(m-1)}{m},\nonumber\\
&\overline{\mu}=\underline{\mu}+\frac{1}{m}.
\end{align} 
According to \cite{kamgar-parsi1995distribution}, the probability density function of $\mu$, i.e., $f_{\mu}(u)$, is symmetric about its expected value $E(\mu)$. Thus, it holds that
\begin{align}\label{eq:EmuApu}
&E(\mu)=\frac{\underline{\mu}+\overline{\mu}}{2}=\frac{(1-w_i)m+w_i-1/2}{m},\nonumber \\
&f_{\mu}(E(\mu)-x)=f_{\mu}(E(\mu)+x)~\forall~x\in \mathbb{R}.
\end{align}
It is next shown that when $E(\mu)=(k-1)/m \Leftrightarrow w_i= (m-k+1/2)/(m-1)$, the probabilities of the states $x_C^k$ and $x_C^{k-1}$ of the child node $X_C$ generated with WMEAN for the scenario $x^{D,i}$ become roughly equal. This result is established by first deriving expressions for the bounds of the difference of the probabilities and then exploring the bounds numerically. Correspondingly, it is shown that when $E(\mu)=(k-1/2)/m \Leftrightarrow w_i= (m-k)/(m-1)$, the probabilities of the states $x_C^{k-1}$ and $x_C^{k+1}$ of the child node $X_C$ generated with WMEAN for the scenario $x^{D,i}$ become roughly equal. 

As a common starting point regarding both of the $E(\mu)$ cases discussed above, consider the difference of the probabilities of the states $x_C^{k_1}$ and $x_C^{k_2}$ of the child node. By using the substitutions 
\begin{align}\label{eq:p1andp2}
&p_1(u)=\int_{\frac{k_1-1}{m}}^{\frac{k_1}{m}}Normpdf(x,u,\sigma^2)dx,\nonumber\\
&p_2(u)=\int_{\frac{k_2-1}{m}}^{\frac{k_2}{m}}Normpdf(x,u,\sigma^2)dx,\nonumber\\
&A(u)=\int_0^1 Normpdf(x,u,\sigma^2)dx,
\end{align}
it is obtained on the basis of Eqs. \eqref{eq:WMEANDscen} and \eqref{eq:EmuApu} that
\begin{align}
&P(X_C=x_C^{k_1}|x^{D,i}; \WMEAN,(w_1,\ldots,w_n),\sigma^2,s)\nonumber\\
&-P(X_C=x_C^{k_2}|x^{D,i}; \WMEAN,(w_1,\ldots,w_n),\sigma^2,s)\xrightarrow{s\rightarrow \infty} \nonumber\\
&\int_{\underline{\mu}}^{\overline{\mu}}\left[\int_{\frac{k_1-1}{m}}^{\frac{k_1}{m}}TNormpdf(x,u,\sigma^2,0,1)dx\right] f_{\mu}(u)du\nonumber\\
\addlinespace
&-\int_{\underline{\mu}}^{\overline{\mu}}\left[\int_{\frac{k_2-1}{m}}^{\frac{k_2}{m}}TNormpdf(x,u,\sigma^2,0,1)dx\right] f_{\mu}(u)du\nonumber
\end{align}
\begin{align}\label{eq:probaEro}
&=\int_{\underline{\mu}}^{\overline{\mu}}\frac{p_1(u)-p_2(u)}{A(u)}f_{\mu}(u)du\nonumber\\
\addlinespace
&=\int_{\underline{\mu}}^{E(\mu)}\frac{p_1(u)-p_2(u)}{A(u)}f_{\mu}(u)du+\int_{E(\mu)}^{\overline{\mu}}\frac{p_1(u)-p_2(u)}{A(u)}f_{\mu}(u)du\nonumber\\
\addlinespace
&=\int_{0}^{\frac{\overline{\mu}-\underline{\mu}}{2}}\frac{p_1(E(\mu)-y)-p_2(E(\mu)-y)}{A(E(\mu)-y)}f_{\mu}(E(\mu)-y)dy\nonumber\\
\addlinespace
&+\int_{0}^{\frac{\overline{\mu}-\underline{\mu}}{2}}\frac{p_1(E(\mu)+y)-p_2(E(\mu)+y)}{A(E(\mu)+y)}f_{\mu}(E(\mu)+y)dy\nonumber\\
\addlinespace
&=\int_{0}^{\frac{\overline{\mu}-\underline{\mu}}{2}}\left[\frac{p_1(E(\mu)-y)-p_2(E(\mu)-y)}{A(E(\mu)-y)}+\frac{p_1(E(\mu)+y)-p_2(E(\mu)+y)}{A(E(\mu)+y)}\right]\nonumber\\
\addlinespace
&*f_{\mu}(E(\mu)+y)dy.
\end{align} 

Next, in Section \ref{subsubsec:BoundDer}, the expression for the probability difference derived in Eq. \eqref{eq:probaEro} is utilized to derive bounds for the difference when $E(\mu)=(k-1)/m$, $k_1=k-1$, and $k_2=k$. Then, in Section \ref{subsubsec:numStudy}, these bounds are numerically studied for further insight. After that, Section \ref{subsubsec:BoundDer2} presents the results of corresponding derivations and numerical studying when $E(\mu)=(k-1/2)/m$, $k_1=k-1$, and $k_2=k+1$. 

\subsection{Bounds for probability difference in Eq. \eqref{eq:probaEro} when $E(\mu)=(k-1)/m$, $k_1=k-1$ and $k_2=k$}\label{subsubsec:BoundDer}

Let
\begin{equation}\label{eq:ehdot}
\left\{\begin{array}{lr}
E(\mu)=\frac{k-1}{m}<\frac{1}{2} \Leftrightarrow w_i= \frac{m-k+1/2}{m-1}~~\mbox{and}~~ k<\frac{m}{2}+1\\
k_1=k-1 \\
k_2=k
\end{array}\right..
\end{equation}
Then, it follows from Eqs. \eqref{eq:muBounds3} and \eqref{eq:p1andp2} that
\begin{align}\label{eq:p1p2}
&\underline{\mu}=\frac{k-3/2}{m},\quad \overline{\mu}=\frac{k-1/2}{m},\quad \frac{\overline{\mu}-\underline{\mu}}{2}=\frac{1/2}{m},\nonumber\\
&p_1(E(\mu)-y)=p_2(E(\mu)+y)~~\forall~y\in \left[-\tfrac{1/2}{m},\tfrac{1/2}{m}\right],
\end{align}
whereby Eq. \eqref{eq:probaEro} can be written as 
\begin{align}
&P(X_C=x_C^{k-1}|x^{D,i}; \WMEAN,(w_1,\ldots,w_n),\sigma^2,s)\nonumber\\
&-P(X_C=x_C^k|x^{D,i}; \WMEAN,(w_1,\ldots,w_n),\sigma^2,s)\xrightarrow{s\rightarrow \infty} \nonumber\\
&\int_{0}^{\frac{1/2}{m}}\left(p_2(\tfrac{k-1}{m}+y)-p_1(\tfrac{k-1}{m}+y)\right)\left[\frac{1}{A(\tfrac{k-1}{m}-y)}-\frac{1}{A(\tfrac{k-1}{m}+y)}\right]f_{\mu}(\tfrac{k-1}{m}+y)dy\nonumber\\
&=\int_{0}^{\frac{1/2}{m}}h(y)f_{\mu}(\tfrac{k-1}{m}+y)dy,
\end{align}
where $h(y)$ is defined as
\begin{align}
h(y)=\left(p_2(\tfrac{k-1}{m}+y)-p_1(\tfrac{k-1}{m}+y)\right)\left[\frac{1}{A(\tfrac{k-1}{m}-y)}-\frac{1}{A(\tfrac{k-1}{m}+y)}\right].
\end{align}
Note that $h(y)$ is non-negative and increasing when $y\in[0,\frac{1/2}{m}]$. This stems from the fact that both of the differences $p_2(\tfrac{k-1}{m}+y)-p_1(\tfrac{k-1}{m}+y)$ and $\frac{1}{A(\tfrac{k-1}{m}-y)}-\frac{1}{A(\tfrac{k-1}{m}+y)}$ are non-negative and increasing when $y\in[0,\frac{1/2}{m}]$. Thus, it applies 
\begin{equation}\label{eq:probaerononzero}
\int_{0}^{\frac{1/2}{m}}h(y)f_{\mu}\left(\tfrac{k-1}{m}+y\right)dy>0,
\end{equation}
which provides a lower bound for the difference of the probabilities in Eq. \eqref{eq:probaEro}.

Based on the symmetry of $f_{\mu}(u)$ around $E(\mu)=\frac{k-1}{m}$, it holds that 
\begin{equation}\label{eq:IntHalf}
\int_{0}^{\frac{1/2}{m}}f_{\mu}\left(\tfrac{k-1}{m}+y\right)dy=\frac{1}{2}=\int_{0}^{\frac{1/2}{m}}m*dy.
\end{equation}

Moreover, as $f_{\mu}(\tfrac{k-1}{m}+y)$ is a decreasing function when $y\in(0,\frac{1/2}{m}]$ \cite{kamgar-parsi1995distribution} and because $f_{\mu}(\frac{k-1/2}{m})=0$, there must be a point $y^*\in(0,\frac{1/2}{m}]$ such that
\begin{equation}\label{eq:yStar}
\left\{ \begin{array}{l}
 f_{\mu}(\tfrac{k-1}{m}+y)\geq m~\forall~y\in[0,y^*] \\
  f_{\mu}(\tfrac{k-1}{m}+y)< m~\forall~y\in(y^*,\frac{1/2}{m}]
       \end{array} \right..
\end{equation}

By applying Eqs. \eqref{eq:IntHalf} and \eqref{eq:yStar}, one obtains 
\begin{align}
&\int_{0}^{\frac{1/2}{m}}h(y)f_{\mu}\left(\tfrac{k-1}{m}+y\right)dy- \int_{0}^{\frac{1/2}{m}}m*h(y) dy\nonumber\\
&=\int_{0}^{y^*}h(y)\left(f_{\mu}\left(\tfrac{k-1}{m}+y\right)-m\right)dy\nonumber\\
&+\int_{y^*}^{\frac{1/2}{m}}h(y)\left(f_{\mu}\left(\tfrac{k-1}{m}+y\right)-m\right)dy\nonumber\\
&\leq h(y^*)\int_{0}^{y^*}\left(f_{\mu}\left(\tfrac{k-1}{m}+y\right)-m\right)dy\nonumber\\
&+h(y^*)\int_{y^*}^{\frac{1/2}{m}}\left(f_{\mu}\left(\tfrac{k-1}{m}+y\right)-m\right)dy\nonumber
\end{align}
\begin{align}
&=h(y^*)\int_{0}^{\frac{1/2}{m}}\left(f_{\mu}\left(\tfrac{k-1}{m}+y\right)-m\right)dy\nonumber\\
&=0\nonumber\\
&\Leftrightarrow \nonumber\\
&\int_{0}^{\frac{1/2}{m}}h(y)f_{\mu}\left(\tfrac{k-1}{m}+y\right)dy \leq m\int_{0}^{\frac{1/2}{m}}h(y)dy,
\end{align}
which provides an upper bound for the difference of the probabilities in Eq. \eqref{eq:probaEro}.

If the above derivations are repeated so that in Eq. \eqref{eq:ehdot} one defines $E(\mu)>\frac{1}{2}$, the result that follows is 
\begin{equation}
m\int_{0}^{\frac{1/2}{m}}h(y)dy\leq\int_{0}^{\frac{1/2}{m}}h(y)f_{\mu}\left(\tfrac{k-1}{m}+y\right)dy<0.
\end{equation}
On the other hand, if one defines $E(\mu)=\frac{1}{2}$ in Eq. \eqref{eq:ehdot}, $h(y)$ becomes identically equal to zero.

To summarize, the following result has been derived:
\begin{align}\label{eq:probaEroVirhe}
&E(\mu)=\frac{k-1}{m} \Leftrightarrow w_i= \frac{m-k+1/2}{m-1}\nonumber\\
&\Rightarrow\nonumber\\
&P(X_C=x_C^{k-1}|x^{D,i}; \WMEAN,(w_1,\ldots,w_n),\sigma^2,s)\nonumber\\
&-P(X_C=x_C^k|x^{D,i}; \WMEAN,(w_1,\ldots,w_n),\sigma^2,s)\xrightarrow{s\rightarrow \infty} \nonumber\\
&\int_{0}^{\frac{1/2}{m}}h(y)f_{\mu}\left(\tfrac{k-1}{m}+y\right)dy,\nonumber
\end{align} 
\begin{align}
&\left|\int_{0}^{\frac{1/2}{m}}h(y)f_{\mu}\left(\tfrac{k-1}{m}+y\right)dy\right|\leq \left|m\int_{0}^{\frac{1/2}{m}}h(y)dy\right|,
\end{align} 
where 
\begin{align}\label{eq:hFunc}
&h(y)=\nonumber\\
\addlinespace
&=\left[\int_{\frac{k-1}{m}}^{\frac{k}{m}}Normpdf(x,\tfrac{k-1}{m}+y,\sigma^2)dx-\int_{\frac{k-2}{m}}^{\frac{k-1}{m}}Normpdf(x,\tfrac{k-1}{m}+y,\sigma^2)dx\right]\nonumber\\
&*\left[\frac{1}{\int_{0}^{1}Normpdf(x,\frac{k-1}{m}-y,\sigma^2)dx}-\frac{1}{\int_{0}^{1}Normpdf(x,\frac{k-1}{m}+y,\sigma^2)dx}\right],
\end{align}
and $f_{\mu}(\cdot)$ denotes the probability density function of $\mu$ defined in Eq. \eqref{eq:muDwmean}.

Equation \eqref{eq:probaEroVirhe} provides an expression for an upper bound of the probability difference
\begin{align}\label{probaEroApu}
&P(X_C=x_C^{k-1}|x^{D,i}; \WMEAN,(w_1,\ldots,w_n),\sigma^2,s)\nonumber\\
&-P(X_C=x_C^k|x^{D,i}; \WMEAN,(w_1,\ldots,w_n),\sigma^2,s)
\end{align}
when $w_i= (m-k+1/2)/(m-1)$ and the sample size parameter $s$ is infinitely large. Next, this quantity is numerically studied to get further insight on its magnitude in practice. 

\subsection{Numerical study on upper bound of probability difference in Eq. \eqref{eq:probaEroVirhe}}\label{subsubsec:numStudy}

Concerning the result \eqref{eq:probaEroVirhe}, the functions $D_{ub}$, $D_{RNM5}$, and $D_{RNM10}$ are defined as
\begin{align}\label{eq:errorMeasures}
&D_{ub}(m,k,\sigma^2)=\left|m\int_{0}^{\frac{1/2}{m}}h(y)dy\right|,\nonumber\\
\addlinespace
&D_{RNM5}(n,m,k,\boldsymbol{w},\sigma^2)=\left|P(X_C=x_C^{k-1}|x^{D,i}; \WMEAN,\boldsymbol{w},\sigma^2,s=5)\right.\nonumber\\
&\quad\qquad\qquad\qquad\qquad\qquad\left.-P(X_C=x_C^{k}|x^{D,i}; \WMEAN,\boldsymbol{w},\sigma^2,s=5)\right|,\nonumber\\
\addlinespace
&D_{RNM10}(n,m,k,\boldsymbol{w},\sigma^2)=\left|P(X_C=x_C^{k-1}|x^{D,i}; \WMEAN,\boldsymbol{w},\sigma^2,s=10)\right.\nonumber\\
&\quad\qquad\qquad\qquad\qquad\qquad\left.-P(X_C=x_C^{k}|x^{D,i}; \WMEAN,\boldsymbol{w},\sigma^2,s=10)\right|,
\end{align}
where the weights $\boldsymbol{w}=(w_1,\ldots,w_n)$ fulfill
\begin{equation}\label{eq:weightCond}
w_i= \frac{m-k+1/2}{m-1},~w_j\in [0,1]~\forall~j=1,\ldots,n,~j\neq i,~\sum_{j=1}^n w_j=1.
\end{equation} 

Now, $D_{ub}$ is an upper bound for the probability difference \eqref{probaEroApu} when $w_i= (m-k+1/2)/(m-1)$ and the sample size parameter $s$ is infinitely large. The functions $D_{RNM5}$ and $D_{RNM10}$ provide values of the same probability difference when the sample size parameter is fixed to $s=5$ and $s=10$, respectively. Note that while some of the arguments of the functions are not explicitly visible in the expressions in Eq. \eqref{eq:errorMeasures}, their presence in the functions is indicated by Eqs. \eqref{eq:WMEANDscen}, \eqref{eq:muDwmean}, and \eqref{eq:hFunc}. 

The maximum values of the functions $D_{ub}$, $D_{RNM5}$, and $D_{RNM10}$ with respect to their arguments are next studied in order to get concrete understanding about the magnitude of the probability difference \eqref{probaEroApu}. Regarding this, based on the derivations resulting in Eq. \eqref{eq:probaEroVirhe}, the following features can be deduced about the functions:
\begin{itemize}
\item Each of the functions fulfills $D_*(n,m,k,\sigma^2)=D_*(n,m,m-k+2,\sigma^2)$ with any $k=2,\ldots,m$.
\item The values of all the three functions decrease with an increasing value of the state index $k$ when $k\leq m/2+1$. Correspondingly, the values of all the functions increase with an increasing value of $k$ when $k>m/2+1$.
\end{itemize} 
In addition, tentative numerical analysis of the functions (results not presented here) indicates the following features:
\begin{itemize}
\item The values of $D_{RNM5}$ and $D_{RNM10}$ decrease with an increasing value of $n$.
\item Given that the weights fulfill Eq. \eqref{eq:weightCond}, the values of the weights other than $w_i$ have little significance to the values of the functions $D_{RNM5}$ and $D_{RNM10}$.
\end{itemize}

\begin{figure*}[!h]
\centering
\subfloat[]{\includegraphics[scale=0.4]{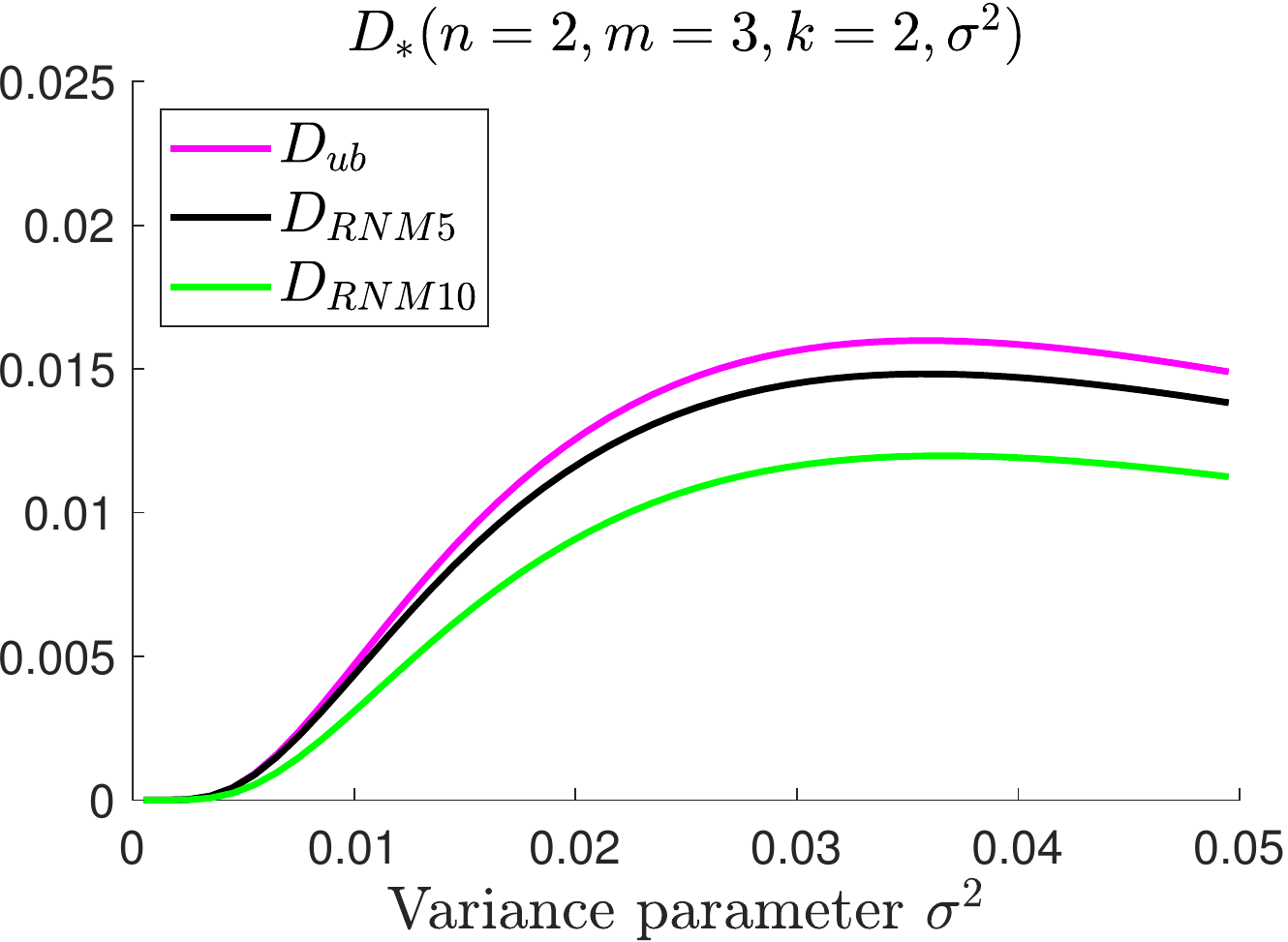}}
~~~
\subfloat[]{\includegraphics[scale=0.4]{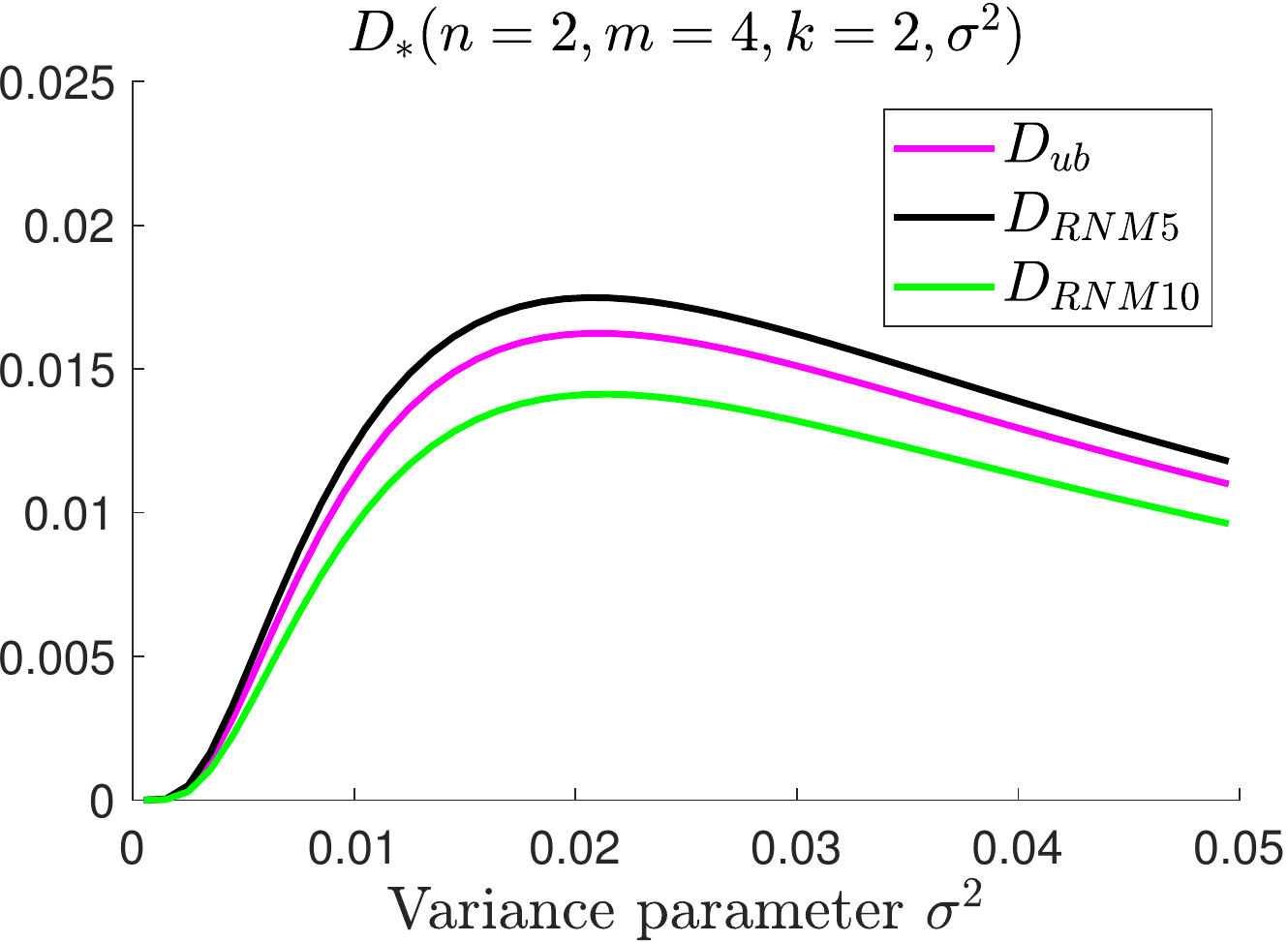}}
~~~
\subfloat[]{\includegraphics[scale=0.4]{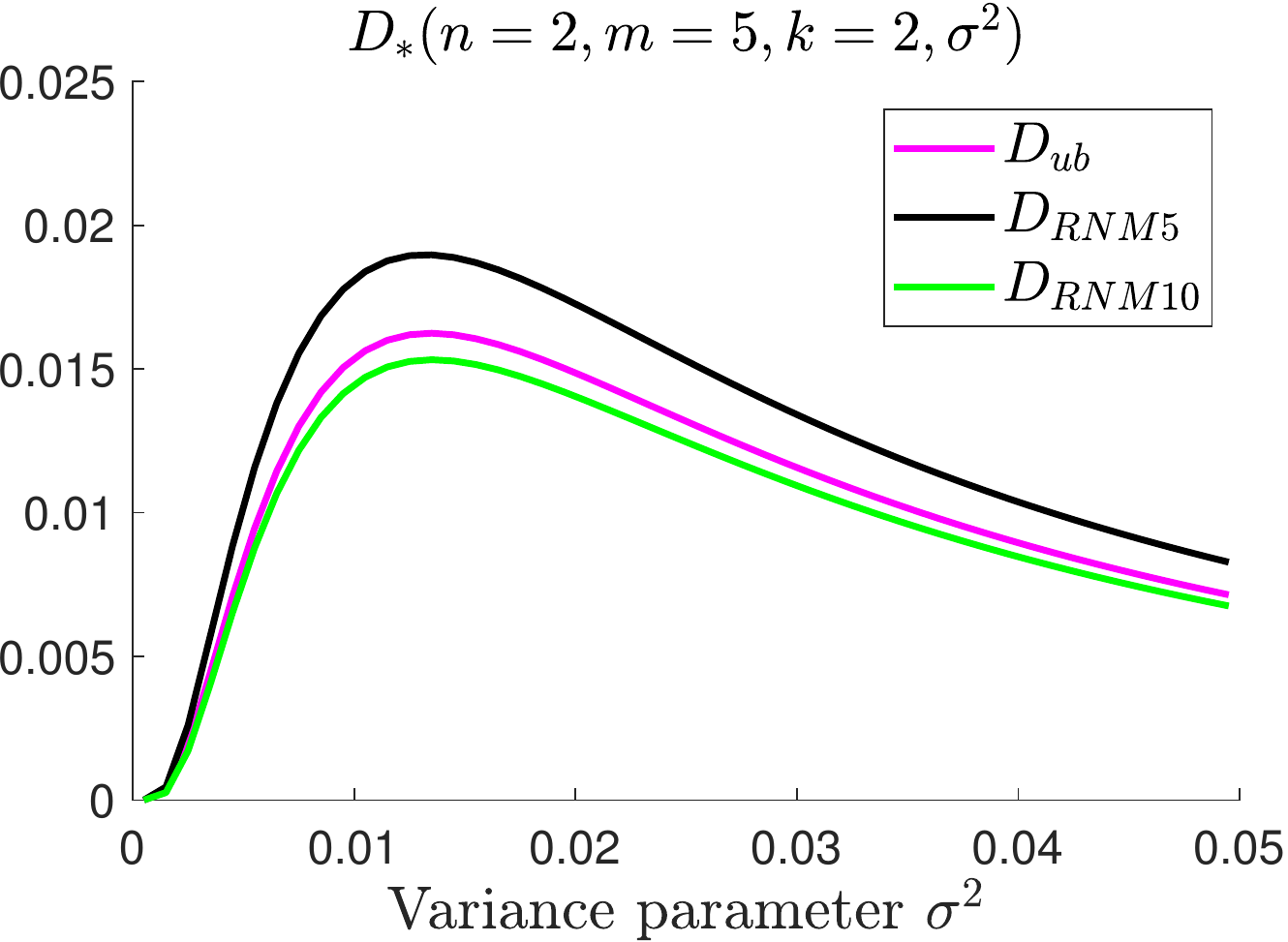}}
\vfill
\subfloat[]{\includegraphics[scale=0.4]{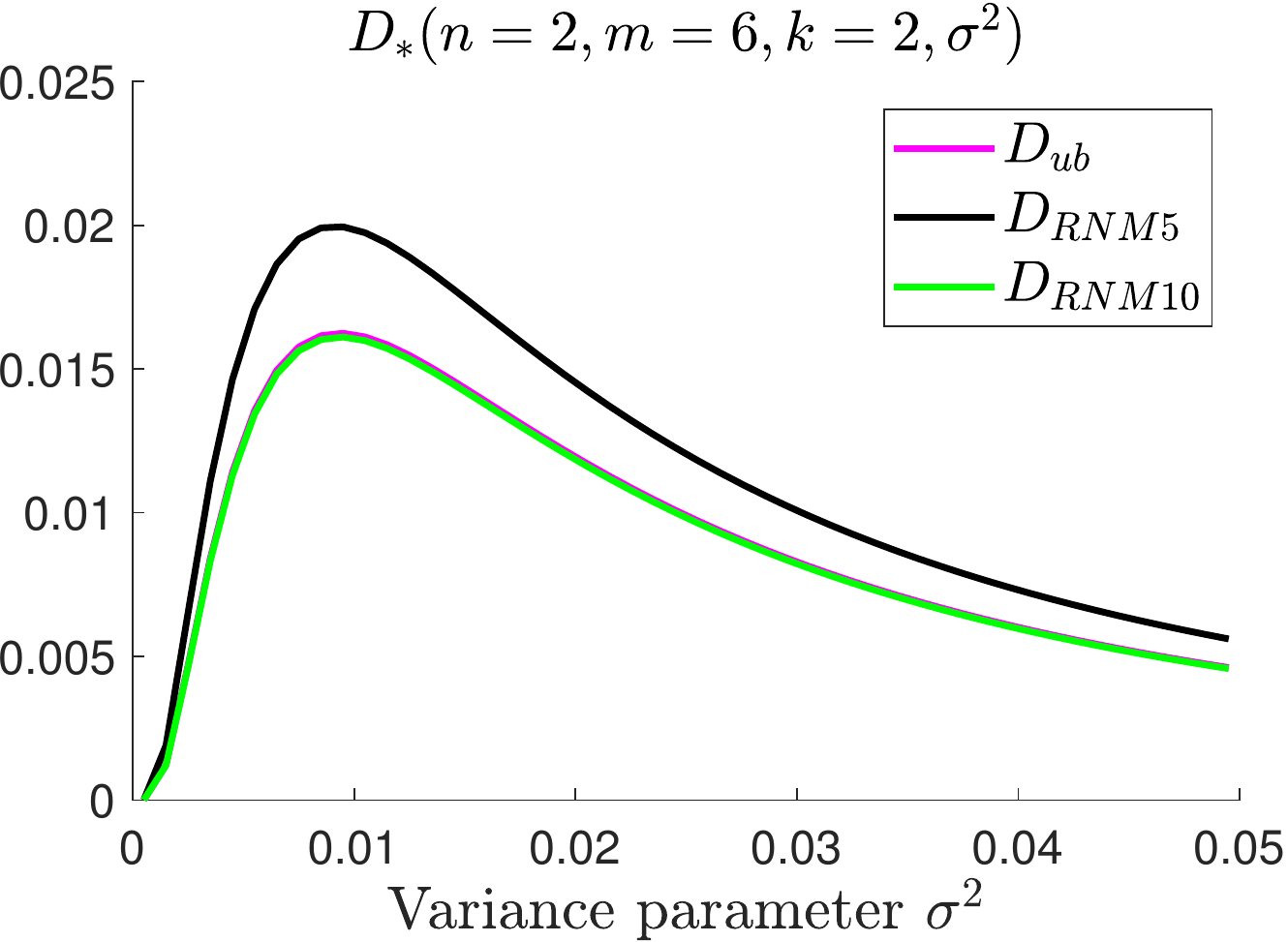}}
~~~
\subfloat[]{\includegraphics[scale=0.4]{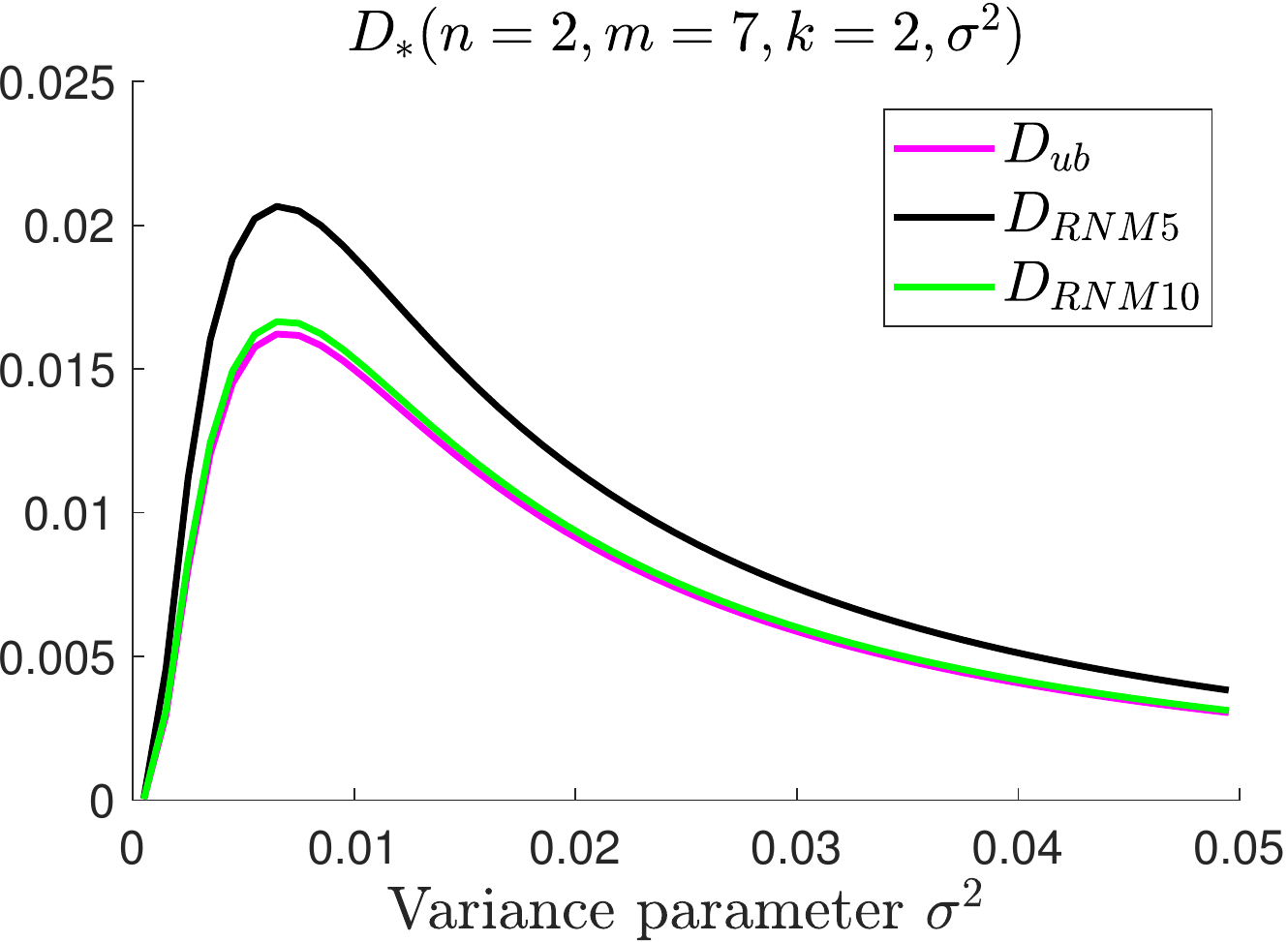}}
~~~
\subfloat[]{\includegraphics[scale=0.4]{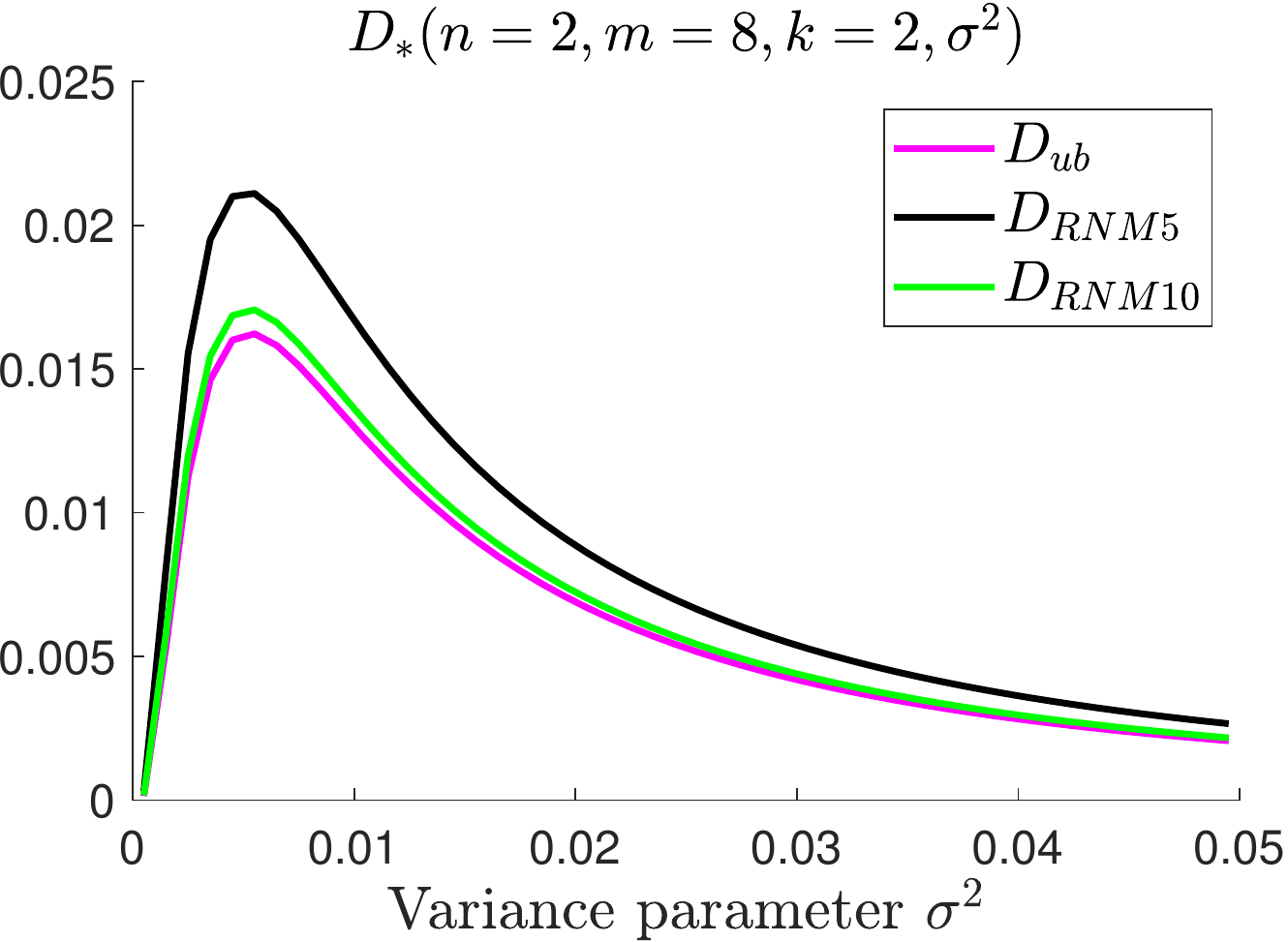}}
\vfill
\subfloat[]{\includegraphics[scale=0.4]{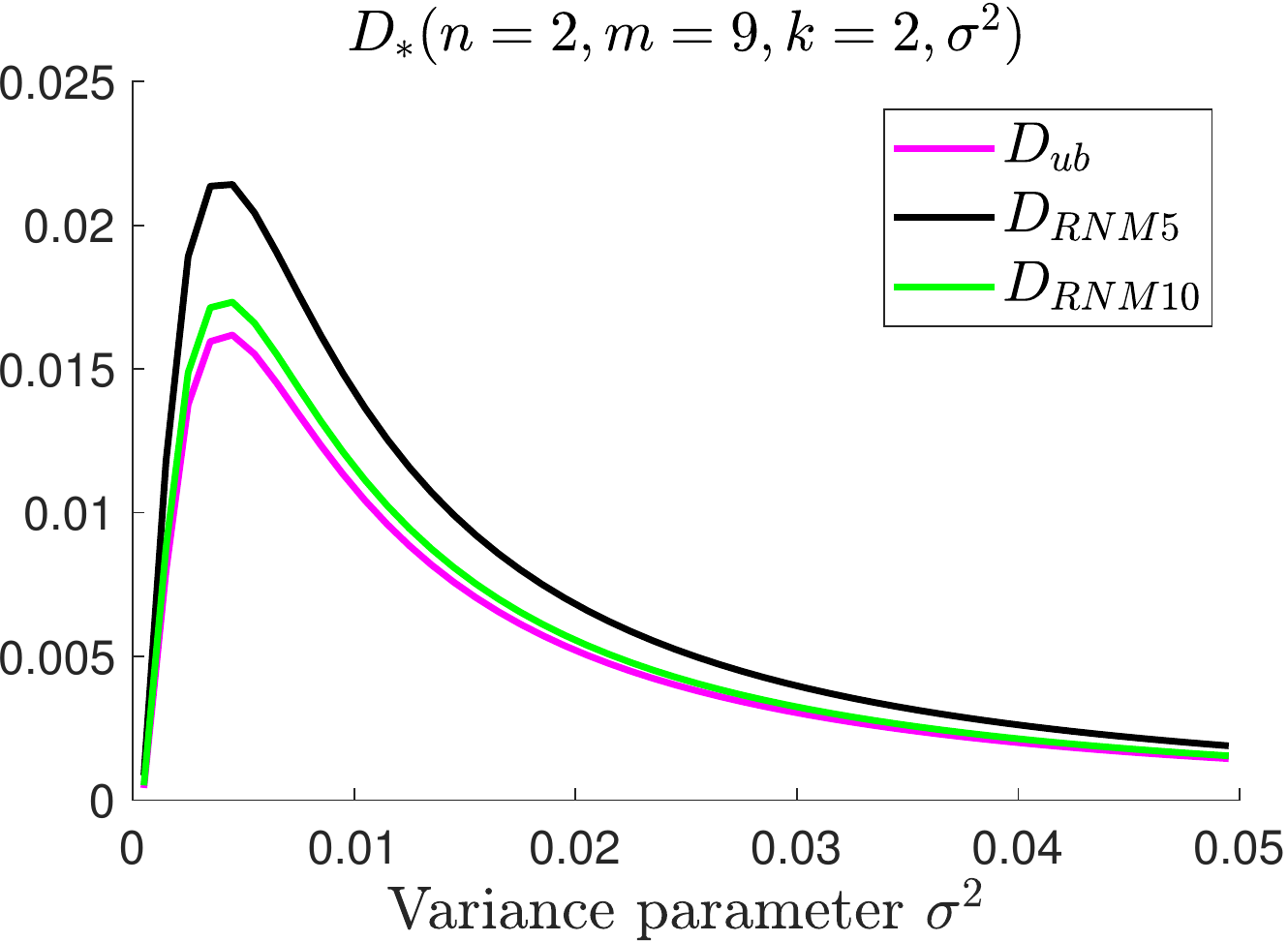}}
~~~
\subfloat[]{\includegraphics[scale=0.4]{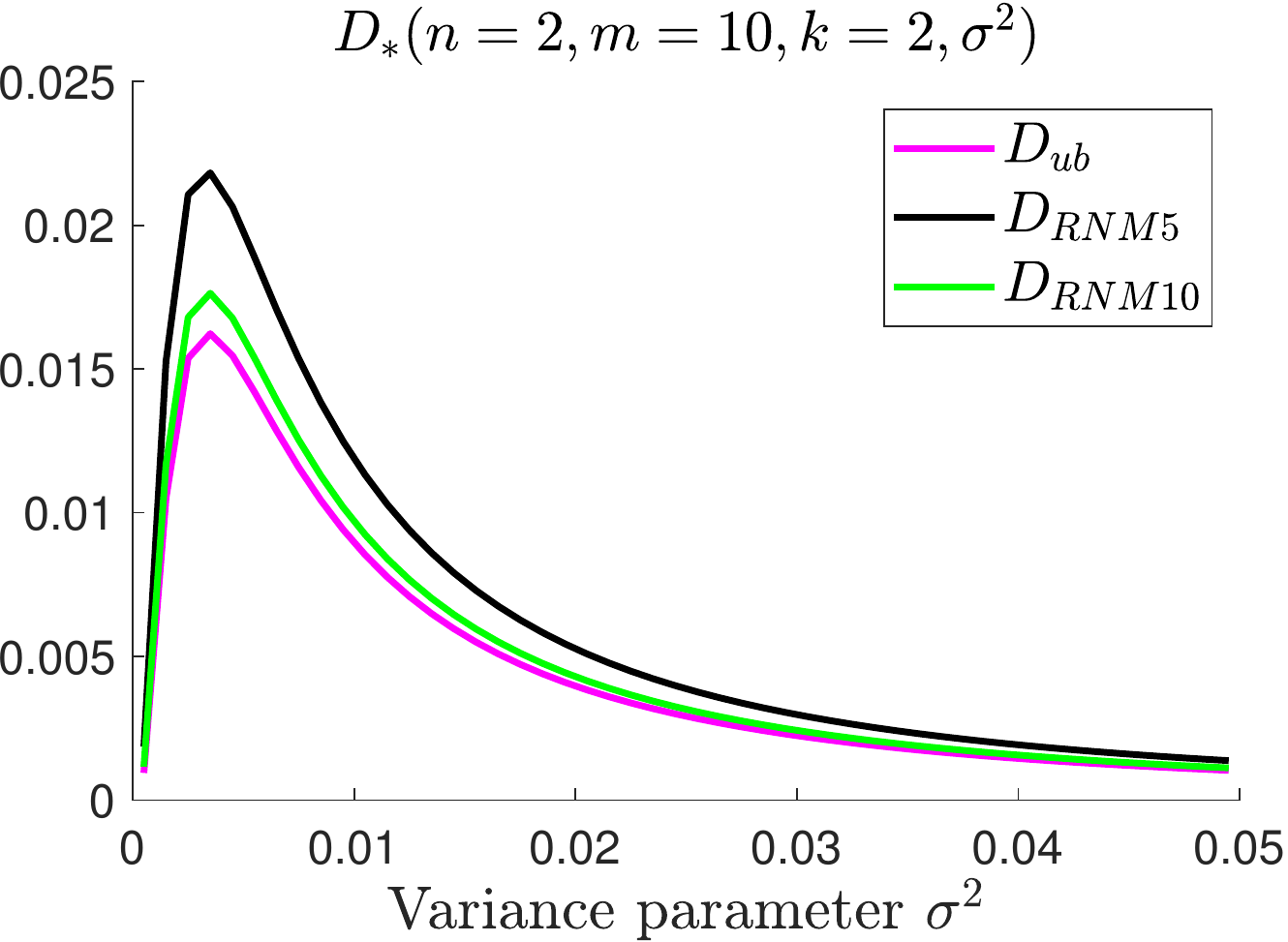}}
~~~
\subfloat[]{\includegraphics[scale=0.4]{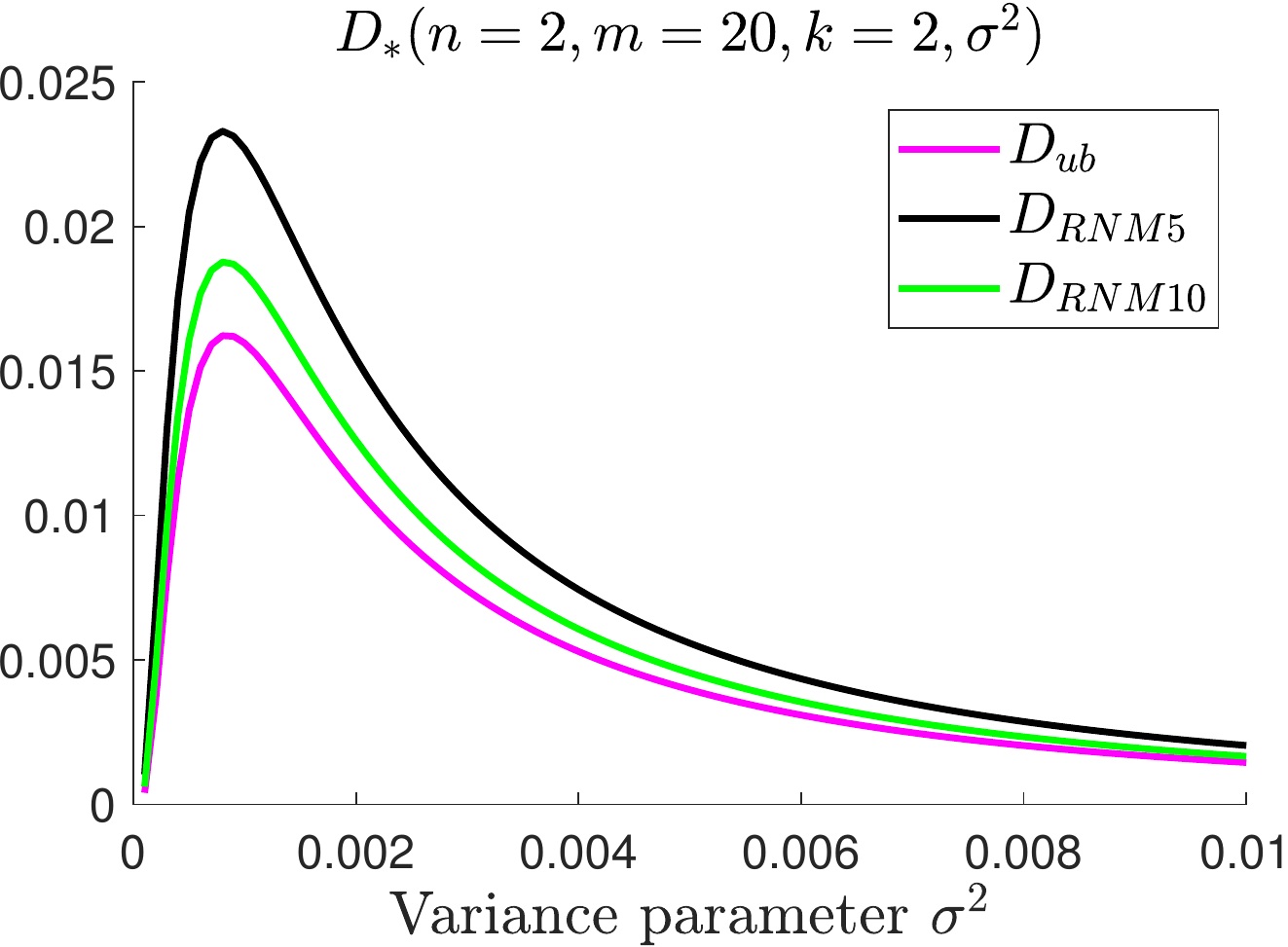}}
\caption{The functions $D_{ub}$, $D_{RNM5}$ and $D_{RNM10}$ plotted as a function of the variance parameter $\sigma^2$ with $n=2$, $k=2$ and (a) $m=3$, (b) $m=4$, (c), $m=5$, (d) $m=6$, (e) $m=7$, (f) $m=8$, (g) $m=9$, (h) $m=10$, and (i) $m=20$. Note the different range of $\sigma^2$ in (i) compared to the other cases.}
\label{fig:D}
\end{figure*} 

The four features listed above indicate that in order to get insight of the overall maximum values of the functions with respect to their arguments, it is sufficient to focus on the cases in which $n=2$ and $k=2$. Now, Fig. \ref{fig:D} presents each of $D_{ub}$, $D_{RNM5}$, and $D_{RNM10}$ as a function of the variance parameter $\sigma^2$ when $n=2$ and $k=2$. The number of the states of the nodes $m$ varies between $m=3$ and $m=20$. Fig. \ref{fig:D} implies that the maximum value of each function with respect to $\sigma^2$ tends to increase with an increasing value of $m$. The value of $D_{ub}$ stays below 0.02 in all of the cases. On the other hand, both $D_{RNM5}$ and $D_{RNM10}$ stay in each case below 0.025. The tendency of $D_{RNM5}$ and $D_{RNM10}$ to be larger than $D_{ub}$ reflects the fact that only a finite number of sample points has been taken while applying RNM in the calculation of the former two. 

Overall, the results presented in Fig. \ref{fig:D} indicate that when WMEAN is used to generate the probability distribution of the scenario $x^{D,i}$, see Eq.  \eqref{eq:sceDi}, with $w_i= \frac{m-k+1/2}{m-1}$ and $s\geq 5$, the difference of the probabilities of the states $x_C^k$ and $x_C^{k-1}$ of the child node are less than 0.025 --- independent of the weights of the other parent nodes or the variance parameter $\sigma^2$ used. That is, it applies
\begin{align}\label{eq:virhe1}
&w_i= \frac{m-k+1/2}{m-1}~~\mbox{and} ~~ s\geq 5\Rightarrow \nonumber\\
&\left|P(X_C=x_C^{k-1}|x^{D,i}; \WMEAN,\boldsymbol{w},\sigma^2,s)- P(X_C=x_C^{k}|x^{D,i}; \WMEAN,\boldsymbol{w},\sigma^2,s)\right| \nonumber\\
&<0.025.
\end{align}  

Based on the proof of Proposition \ref{prop1} in Section \ref{sec:Prop1Proof}, it is evident that $x_C^{k-1}$ and $x_C^k$ are then also the two most probable states of the child node.

\subsection{Results for probability difference in Eq.  \eqref{eq:probaEro} when $E(\mu)=(k-1/2)/m$, $k_1=k-1$ and $k_2=k+1$}\label{subsubsec:BoundDer2}
If the derivations and numerical examinations carried out in Sections \ref{subsubsec:BoundDer} and \ref{subsubsec:numStudy} are repeated so that the conditions \eqref{eq:ehdot} are replaced with  
\begin{equation}\label{eq:ehdot2}
\left\{\begin{array}{lr}
E(\mu)=\frac{k-1/2}{m}<\frac{1}{2} \Leftrightarrow w_i= \frac{m-k}{m-1}~~\mbox{and}~~ k<\frac{m+1}{2}\\
k_1=k-1 \\
k_2=k+1
\end{array}\right.,
\end{equation}
the result corresponding to Eq. \eqref{eq:virhe1} is
\begin{align}\label{eq:virhe2}
&w_i= \frac{m-k}{m-1}~~\mbox{and} ~~ s\geq 5\Rightarrow \nonumber\\
&\left|P(X_C=x_C^{k-1}|x^{D,i}; \WMEAN,\boldsymbol{w},\sigma^2,s)- P(X_C=x_C^{k+1}|x^{D,i}; \WMEAN,\boldsymbol{w},\sigma^2,s)\right| \nonumber\\
&<0.01.
\end{align} 

Equation \eqref{eq:virhe2} implies the value $w_i= \frac{m-k}{m-1}$ is a good approximation of the exact value of $w_i$ by which the states $x_C^{k-1}$ and $x_C^{k+1}$ of the child node become equally probable in the scenario $x^{D,i}$, see Eq. \eqref{eq:sceDi}, when WMEAN is used to generate the probabilities. Moreover, based on the proof of Proposition \ref{prop1} in Section \ref{sec:Prop1Proof}, it is evident that $x_C^{k-1}$ and $x_C^{k+1}$ are then also together the second most probable states of the child node with $x_C^{k}$ being the mode.

Together, Eqs. \eqref{eq:virhe1} and \eqref{eq:virhe2} can be used to determine for the weight $w_i$ bounds $\underline{w}_i^D$ and $\overline{w}_i^D$ such that when $\boldsymbol{w}=(w_1,\ldots,w_n)\in W^{\scriptsize \WMEAN}$ and $w_i \in [\underline{w}_i^D,\overline{w}_i^D]$, the probability distribution $P(X_C|x^{D,i}; \WMEAN,\boldsymbol{w},\sigma^2,s)$ generally has some specific two adjacent states $x_C^a$ and $x_C^b$  of $X_C$ forming its mode pair. Especially, when $w_i =\underline{w}_i^D$ or $w_i=\overline{w}_i^D$, either a)  the mode $x_C^a$ and the second most probable state $x_C^b$ have a probability difference less than 0.025 or b) $x_C^b$ and another state of $X_C$ adjacent to $x_C^a$ have a probability difference smaller than 0.01 between each other.

\section{Experiment on weight updates for WMEAN}\label{sec:WMEANwu}
The results of the previous section indicate that in the scenario $p^{D,i}$, see Eq. \eqref{eq:sceDi}, the weight $w_i$ basically solely defines the mode pair of the child node $X_C$ when the probability distribution $P(X_C|x^{D,i})$ is generated with RNM using WMEAN as the weight expression. Especially, through Eqs. \eqref{eq:virhe1} and \eqref{eq:virhe2}, one can determine an interval $[\underline{w}_i^D,\overline{w}_i^D]$ that provides a good approximation of the values of $w_i$ by which some specific two states of $X_C$ become the mode pair of $P(X_C|x^{D,i})$. Related to this property of WMEAN, this section presents an experiment about the effect of weights other than $w_i$ to the probability values in $P(X_C|x^{D,i})$. The experiment is focused on a specific process of changing weights from some initial values $w_i^0$ to final ones $w_i^*$ where both $w_i^0$ and $w_i^*$ are within the same interval $[\underline{w}_i^D,\overline{w}_i^D]$.

The process of changing the weights is as follows. Let there be initial weights $\boldsymbol{w}^0=(w_1^0,\ldots,w_n^0)\in W^{\scriptsize \WMEAN}$ that are in ascending order, i.e., $w_1^0\leq \ldots \leq w_n^0$, and yield probability distributions $P(X_C|x^{D,i};\boldsymbol{w}^0)$, $i=1,\ldots,n$, with a fixed variance parameter $\sigma^2$. Let $p^{D,i}$ denote the mode pair of $P(X_C|x^{D,i};\boldsymbol{w}^0)$ and $[\underline{w}_i^D,\overline{w}_i^D]$ be the approximation of the values of $w_i$ by which $P(X_C|x^{D,i})$ stays compatible with $p^{D,i}$. Now, starting from $w_n$ and moving towards $w_1$, the initial weights $\boldsymbol{w}^0$ are transformed into the final weights $\boldsymbol{w}^*=(w_1^*,\ldots,w_n^*)\in W^{\scriptsize \WMEAN}$, $w_i^*\in [\underline{w}_i^D,\overline{w}_i^D]$, one by one as follows. When $w_i$ is given the final value $w_i^*$, the weights $w_{i+1},\ldots,w_n$ stay intact in the values $w_{i+1}^*,\ldots,w_n^*$. On the other hand, the weights $w_j$, $j=1,\ldots,i-1$, are updated to any values $\tilde{w}_j\in [\underline{w}_j^D,\overline{w}_j^D]$ such that the intermediate weights $\boldsymbol{w}^i=(\tilde{w}_1,\ldots,\tilde{w}_{i-1},w_i^*,\ldots,w_n^*)$ fulfill $\boldsymbol{w}^i\in W^{\scriptsize \WMEAN}$. The aim in the experiment is to determine, how much there is difference between a probability distribution $P(X_C|x^{D,i};\boldsymbol{w}^i)$ generated with the intermediate weights $\boldsymbol{w}^i$ and $P(X_C|x^{D,i};\boldsymbol{w}^*)$ obtained with the final weights $\boldsymbol{w}^*$. 

\subsection{Experimental setup}
The experiment consists of $N$ replications. On each replication, the values of the following variables are first randomly generated by sampling uniformly from their ranges. 

\begin{itemize}
\item $m\in[3,7],~m\in \mathbb{N}$. The number of states of all the nodes.  The range of $m$ covers the number of states typically given to ranked nodes in practical applications.
\item $n\in[3,8],~n\in \mathbb{N}$. The number of parent nodes.  The range of $n$ covers the number of parent nodes typically defined for a child node in practical applications.
\item  $\tilde{w}_i \in [0,1],~i=1,\ldots,n$. Auxiliary variables used to construct initial weights of parent nodes.
\item $\sigma^2\in[5*10^{-4},\frac{1}{4m^2}]$. The variance parameter that is used in the generation of $P(X_C|x^{D,i})$. The lower bound is the default lower bound for $\sigma^2$ in AgenaRisk software \cite{agenarisk10}  implementing RNM. The rational for the upper bound is explained in the end of this section.
\end{itemize}

The auxiliary variables $\tilde{w}_1,\ldots,\tilde{w}_n$ are used to form initial weights $\boldsymbol{w}^0=(w_1^0,\ldots,w_n^0)\in W^{\scriptsize \WMEAN}$ through
\begin{equation}
w_i^0=\frac{\tilde{w}_i}{\sum_{j=1}^n\tilde{w}_j},
\end{equation}
after which $\boldsymbol{w}^0$ is sorted into ascending order, i.e., $w_1^0\leq \ldots \leq w_n^0$. Next, for $i=1,\ldots,n$, the probability distribution $P(X_C|x^{D,i};\boldsymbol{w}^0)$ is generated using WMEAN with the variance parameter $\sigma^2$ and the interval $[\underline{w}_i^D,\overline{w}_i^D]$ is formed according to the mode pair $p^{D,i}$ of $P(X_C|x^{D,i};\boldsymbol{w}^0)$.

Starting from $w_n$ with the initial value $w_n^0$, the weights $w_i$, $i=2,\ldots,n$, are addressed one by one in a descending order as follows: Given the initial value of the weight $w_i$, two random numbers $u\sim U(0,1)$ and $y\sim U(0,1)$ are generated, and a final value $w_i^*$ for the weight $w_i$ is formed according to
\begin{align}\label{eq:wiUpdate}
&w_i^*=w_i+\Delta_i, \nonumber\\
&\Delta_i=\left\{\begin{array}{l l}
\max\{y*\sum_{j=1}^{i-1}(w_j-\overline{w}_j^D),~\underline{w}_i^D-w_i\}&\qquad\mbox{ if $u< 0.5$}\\\addlinespace
\min\{y*\sum_{j=1}^{i-1}(w_j-\underline{w}_j^D),~\overline{w}_i^D-w_i\}&\qquad\mbox{ if $u\geq 0.5$}
\end{array}\right..
\end{align}
This update scheme means that with equal probabilities, the value of $w_i$ either decreases or increases so that the final value $w_i^*$ is still within the interval $[\underline{w}_i^D,\overline{w}_i^D]$. After updating $w_i$, the values of the weights $w_j$, $j=1,\ldots,i-1$, are updated to new values $\tilde{w}_j$ according to
\begin{align}\label{eq:wjUpdate}
&\tilde{w}_j=w_j+\Delta_j,\nonumber\\
&\Delta_j=\left\{\begin{array}{ll}
-\frac{\Delta_i}{\sum_{t=1}^{i-1}(\overline{w}_t^D-w_t)}(\overline{w}_j^D-w_j)&\qquad\mbox{ if $\Delta_i< 0$}\\\addlinespace
-\frac{\Delta_i}{\sum_{t=1}^{i-1}(w_t-\underline{w}_t^D)}(w_j-\underline{w}_j^D)&\qquad\mbox{ if $\Delta_i\geq 0$}
\end{array}\right..
\end{align}  
If the value of $w_i$ is decreased (increased) in Eq. \eqref{eq:wiUpdate}, the values of the weights $w_1,\ldots,w_{i-1}$ are all increased (decreased) in Eq. \eqref{eq:wjUpdate}. The magnitudes of all the changes $\Delta_1,\ldots,\Delta_{i}$ are defined so that the updated weights fulfil the conditions
\begin{align}
&\tilde{w}_j\in [\underline{w}_j^D,\overline{w}_j^D]~\forall~j=1,\ldots,i-1,\nonumber\\
&\sum_{j=1}^{i-1}\tilde{w}_j+\sum_{j=i}^n w_j^* =1.
\end{align} 
That is, the updated values of the weights $w_j$, $j=1,\ldots,i-1$, are within their elicitation weight intervals and all the weights sum up to 1 in accordance with Eq. \eqref{eq:wmean}. By using the intermediate weights $\boldsymbol{w}^i=(\tilde{w}_1,\ldots,\tilde{w}_{i-1},w_i^*,\ldots,w_n^*)$, the probability distribution $P(X_C|x^{D,i};\boldsymbol{w}^i)$ is generated and put aside. The sample size of RNM used in the generation is $s=5$ which is the default value of $s$ in AgenaRisk. 

After the final weights $\boldsymbol{w}^f=(w_1^*,\ldots,w_n^*)$ are formed, they are used to generate probability distributions $P(X_C|x^{D,i};\boldsymbol{w}^*)$, $i=3,\ldots,n$. For each $i=3,\ldots,n$, the elements of the probability distribution $P(X_C|x^{D,i};\tilde{\boldsymbol{w}}^*)$ are then compared to those of $P(X_C|x^{D,i};\boldsymbol{w}^i)$. Based on these comparisons, the quantities $e_1$ and $e_2$ are calculated according to
\begin{align}
&e_1=\max_{i=3,\ldots,n}\left\{\frac{1}{m}\sum_{k=1}^m|P(X_C=x_C^k|x^{D,i};\boldsymbol{w}^i)-P(X_C=x_C^k|x^{D,i};\boldsymbol{w}^*)|\right\},\nonumber\\
&e_2=\max_{i=3,\ldots,n}\left\{\max_{k=1,\ldots,m}\{|P(X_C=x_C^k|x^{D,i};\boldsymbol{w}^i)-P(X_C=x_C^k|x^{D,i};\boldsymbol{w}^*)|\}\right\}.
\end{align}
Here, $e_1$ is the largest average difference observed when the corresponding elements of $P(X_C|x^{D,i};\boldsymbol{w}^*)$ and $P(X_C|x^{D,i};\boldsymbol{w}^i)$ are compared to each other for all $i=3,\ldots,n$. In turn, $e_2$ is the single largest difference observed when the same comparisons are made.   

Let $e_1^r$ and $e_2^r$ denote the values of $e_1$ and $e_2$ obtained in the $r$th replication of the experiment. After all the $N$ replications, the quantities $\bar{e}_1$, $\bar{e}_2$, $\hat{e}_1$, and $\hat{e}_2$ are calculated according to
\begin{align}
&\bar{e}_1=\frac{1}{N}\sum_{r=1}^N e_1^r,
\qquad\bar{e}_2=\frac{1}{N}\sum_{r=1}^N e_2^r\nonumber\\
&\hat{e}_1=\max_{r=1,\ldots,N}\{e_1^r\},
\qquad\hat{e}_2=\max_{r=1,\ldots,N}\{e_2^r\}.
\end{align}
Now, $\bar{e}_1$ is an estimate of the largest average change of probability values that could be expected when the elements of the probability distributions $P(X_C|x^{D,i};\boldsymbol{w}^i)$, $i=3,\ldots,n$, are compared to the elements of the distributions  $P(X_C|x^{D,i};\boldsymbol{w}^*)$. Correspondingly, $\bar{e}_2$ is an estimate of the largest change in a single probability value that could be expected to occur in such a comparison. In turn, $\hat{e}_1$ and $\hat{e}_2$ are worst-case estimates of the same quantities as $\bar{e}_1$ and $\bar{e}_2$.

The upper bound for the variance parameter $\sigma^2$ used in the experiment is $1/m^2$. When the number of states of the nodes $m$ runs from 3 to 7,  the value of $1/m^2$ runs from ca. 0.1 to ca. 0.02. The default upper bound for $\sigma^2$ in AgenaRisk is 0.5. However, tentative runs of the experiment (results not presented here) indicate that with the upper bound of $\sigma^2$ set this large, there constantly comes replications in which both $P(X_C|x^{D,i};\boldsymbol{w}^*)$ and $P(X_C|x^{D,i};\boldsymbol{w}^i)$ are so close to uniform distributions that their differences become minimal. As probability distributions generated with RNM in practical applications are seldom uniform, a smaller upper bound than 0.5 for $\sigma^2$ is seen more adequate now. The idea for using $\sigma^2=1/m^2$ as the upper bound is as follows. In general,  95\% of the probability mass of a normal distribution $N(\mu,\sigma^2)$ lies within the interval $[\mu-2\sigma,\mu+2\sigma]$. Thus,  
if $\mu=\frac{k}{m}$ and one desires  
\begin{align}\label{eq:int095}
\int_{\frac{k-2}{m}}^{\frac{k+2}{m}}Normpdf(x,\mu,\sigma^2)=0.95,
\end{align}
it must be that $2/m=2\sigma \Leftrightarrow \sigma^2=1/m^2$. Next, consider the generation of probabilities with RNM as described in Section \ref{sec:CPDGen}. First note that when all the $n$ parent nodes have $m$ states, all the mean parameters $\mu_r$, $r=1,\ldots,s^n$, see Eq. \eqref{eq:muK}, necessarily lie within an interval of the width $1/m$. Suppose then that a given mean parameter $\mu_r$ lies within the state interval $[a_C^k,b_C^k]=[\frac{k-1}{m},\frac{k}{m}]$ of the child node. Then, with $2\sigma=2/m$, more than 95\% of the probability mass of the doubly truncated normal distribution $TN(\mu_r,\sigma^2,0,1)$ lies within the state interval $[a_C^k,b_C^k]$ and its three closest neighbouring state intervals which depend on the exact value of $\mu_r$. This fact and the limited range of the mean parameters $\mu_r$, $r=1,\ldots,s^n$, together imply that if $\sigma^2=1/m^2$ is used as the variance parameter, the majority of the probability mass in a given generated conditional probability distribution $P(X_C|x^{D,i})$ is shared between the four or the five most probable states of the child node. This type of spread of the probability mass of $X_C$ is now considered to represent well a heavy dispersion. 

\subsection{Results}\label{subsec:ResultsExperimentWMEAN}   

\captionsetup{font={small}}
\begin{table*}[!h]
\caption{Values of $\bar{e}_1$, $\bar{e}_2$, $\hat{e}_1$, and $\hat{e}_2$ obtained with $N=10000$ replications}
\begin{center}
\begin{tabular}{c c c c}
\hline
$\bar{e}_1$ & $\bar{e}_2$ & $\hat{e}_1$ & $\hat{e}_2$ \\
\hline
$0.0006$ & $0.0011$ & $0.019$ & $0.028$\\
\hline
\end{tabular}
\label{tab:ExpWMEANresults}
\end{center}
\end{table*}

Table \ref{tab:ExpWMEANresults} presents the values of $\bar{e}_1$, $\bar{e}_2$, $\hat{e}_1$, and $\hat{e}_2$ obtained in the experiment with $N=10000$ replications. The estimates $\bar{e}_1$ and $\bar{e}_2$ are smaller than $0.002$ while $\hat{e}_1$ and $\hat{e}_2$ both remain under $0.03$. Therefore, the results indicate that there is no considerable difference in the probability distributions $P(X_C|x^{D,i};\boldsymbol{w}^i)$ and $P(X_C|x^{D,i};\boldsymbol{w}^*)$ generated with the intermediate weights $\boldsymbol{w}^i$ and the final weights $\boldsymbol{w}^*$, respectively. 

The results of the experiment may be utilized, e.g., as follows. Let there be initial weights $\boldsymbol{w}^0=(w_1^0,\ldots,w_n^0)\in W^{\scriptsize \WMEAN}$ such that $w_i^0\in [\underline{w}_i^D,\overline{w}_i^D]$ and $w_1^0\leq \ldots \leq w_n^0$. Starting from $w_n$ and moving towards $w_1$, a desired value $w_i^*\in [\underline{w}_i^D,\overline{w}_i^D]$ for the weight $w_i$ can be determined by repeating the following two steps: 1) construct intermediate weights of the form $\boldsymbol{w}^i=(\tilde{w}_1,\ldots,\tilde{w}_{i-1},w_i^*,w_{i+1}^*,\linebreak\ldots,w_n^*)$, where $w_i^*$ is a candidate for the desired value of $w_i$, $w_{i+1}^*,\ldots,w_n^*$ are the already determined desired values of $w_{i+1},\ldots,w_n$, and $\tilde{w}_j$, $j=1\ldots,i-1$, are any values of $w_j\in[\underline{w}_j^D,\overline{w}_j^D]$ so that $\boldsymbol{w}^i\in W^{\scriptsize \WMEAN}$. 2) generate the probability distribution $P(X_C|x^{D,i};\boldsymbol{w}^i)$ and return to Step 1 if the result is not satisfactory. That is, the desired value of $w_i$ is determined through trial and error by checking how different values of $w_i$ affect $P(X_C|x^{D,i},\boldsymbol{w}^i)$. Once desired values $\boldsymbol{w}^*=(w_1^*,\ldots,w_n^*)$ are determined for all the weights, the probability distribution $P(X_C|x^{D,i};\boldsymbol{w}^*)$ generated with them does not differ much from the distribution $P(X_C|x^{D,i};\boldsymbol{w}^i)$ originally accepted while deciding the desired value of $w_i$.

\section{Connection between WMIN and WMEAN}\label{sec:WMINProp}

In this section it is shown that under specific conditions, the functional form of WMIN in the scenario $x^{D,i}$, see Eq. \eqref{eq:sceDi}, coincides with that of WMEAN. Thereby, under these conditions, the technical results obtained for WMEAN in Section \ref{sec:WMEANProps} are applicable also with WMIN.

When the conditional probability distribution $P(X_C|x^{D,i})$ is generated with RNM using WMIN with the weights $(w_1,\ldots,w_n)\in W^{\mbox{\scriptsize WMIN}}$ and the variance parameter $\sigma^2$, it applies, see Eq. \eqref{eq:RNMprincip2},
\begin{align}\label{eq:WMINConv}
&P(X_C=x_C^k|x^{D,i}; \WMIN,(w_1,\ldots,w_n),\sigma^2,s)\xrightarrow{s\rightarrow \infty} \nonumber\\
&\int_{\underline{\mu}}^{\overline{\mu}}\left[\int_{\frac{k-1}{m}}^{\frac{k}{m}}TNormpdf(x,u,\sigma^2,0,1)dx\right] f_{\mu}(u)du,
\end{align}
where $x_C^k$ denotes the $k$th state of the child node $X_C$ and $\mu$ defined in Eq. \eqref{eq:mu} has the form
\begin{align}\label{eq:WMINConvMu}
&\mu=\min\limits_{j=1,\ldots,n}\left\{\frac{w_j \chi_j+\sum_{t\neq j}\chi_t}{w_j+n-1}\right\},~~\chi_i\sim U(0,\tfrac{1}{m}),~\chi_r\sim U(\tfrac{m-1}{m},1)~\forall~r=1,\ldots,n,~r\neq i.
\end{align}
The values of $\underline{\mu}$ and $\overline{\mu}$, i.e., the lower and upper bounds of $\mu$
are now 
\begin{align}\label{eq:WMINConvMuBounds}
&\underline{\mu}=\min\left\{\min\limits_{j\neq i}\left\{\frac{w_j \frac{m-1}{m}+(n-2)\frac{m-1}{m}}{w_j+n-1}\right\},\frac{(n-1)\frac{m-1}{m}}{w_i+n-1}\right\}=\frac{(n-1)\frac{m-1}{m}}{w_i+n-1},\nonumber\\
\addlinespace
&\overline{\mu}=\underline{\mu}+\frac{1}{m}.
.
\end{align} 

By defining random variables $\mu_j$, $j=1,\ldots,n$, as
\begin{equation}\label{eq:mujExpres}
\mu_j(w_j)=\frac{w_j \chi_j+\sum_{t\neq j}\chi_t}{w_j+n-1},~~\chi_i\sim U(0,\tfrac{1}{m}),~\chi_r\sim U(\tfrac{m-1}{m},1)~\forall~r=1,\ldots,n,~r\neq i,
\end{equation}
the variable $\mu$ in Eq. \eqref{eq:WMINConvMu} can be expressed as
\begin{equation}\label{eq:muExpres}
\mu=\min_{j=1,\ldots,n}\{\mu_j(w_j)\}
\end{equation}
Below, Proposition \ref{prop2} states conditions under which $\mu$ is identically equal to $\mu_i$, i.e., $\mu\equiv\mu_i$. When this equation holds, the functional form of WMIN in Eq. \eqref{eq:wmin} coincides with the functional form of WMEAN in Eq. \eqref{eq:wmean}.

\setcounter{proposition}{1}
\begin{proposition}\label{prop2}
Let $\chi_i\sim U(0,\tfrac{1}{m})$ and $\chi_r\sim U(\tfrac{m-1}{m},1)~\forall~r=1,\ldots,n,~r\neq i$. Furthermore, with any $(w_1,\ldots,w_n)\in W^{\mbox{\scriptsize WMIN}}$, let $\mu_1,\ldots,\mu_n$ and $\mu$ be defined by
\begin{align}\label{eq:muexpres}
&\mu_j(w_j)=\frac{w_j \chi_j+\sum_{t\neq j}\chi_t}{w_j+n-1},\nonumber\\
&\mu(w_1,\ldots,w_n)=\min\limits_{j=1,\ldots,n}\left\{\mu_j(w_j)\right\}.
\end{align}
It applies
\begin{equation}\label{eq:prop2aa}
n\leq m \Rightarrow \mu(w_1,\ldots,w_n)\equiv\mu_i(w_i)~\forall~(w_1,\ldots,w_n)\in W^{\mbox{\scriptsize WMIN}},
\end{equation}
\begin{equation}\label{eq:prop2bb}
n>m \Rightarrow \left(\mu(w_1,\ldots,w_n)\equiv\mu_i(w_i)~\forall~w_j\geq 1,~j=1\ldots,n,~j\neq i \Leftrightarrow w_i\geq \frac{n-2}{m-2}\right).
\end{equation}
\end{proposition}
\begin{proof}
The proof is presented below after a brief discussion on the implications of the proposition.
\end{proof}
Now, Eq. \eqref{eq:prop2aa} indicates that when $n\leq m$ (the number of parent nodes is less or equal to the number of states of the nodes), $\mu$ is identically equal to $\mu_i$ with any weights $(w_1,\ldots,w_n)\in W^{\mbox{\scriptsize WMIN}}$. On the other hand, Eq. \eqref{eq:prop2bb} indicates that when $n> m$, the condition $w_i\geq \frac{n-2}{m-2}$ is both necessary and sufficient for $\mu$ being identically equal to $\mu_i$, no matter what values $w_j\geq 1$ the other weights $w_j,~j=1,\ldots,n,~j\neq i$, have. When the condition $\mu\equiv\mu_i$ holds, the variable $\mu$ in Eq. \eqref{eq:WMINConvMu} can be expressed as
\begin{align}\label{eq:wminOnwmean}
&\mu=\frac{w_i \chi_i+\sum_{t\neq i}\chi_t}{w_i+n-1}=\sum_{t=1}^n \beta_t \chi_t,\nonumber\\
&\beta_i=\frac{w_i}{w_i+n-1},\nonumber\\
&\beta_t=\frac{1}{w_i+n-1}~\forall~t=1,\ldots,n,~t\neq i,
\end{align}
whereby $\mu$ is formed by WMEAN defined in Eq.  \eqref{eq:wmean}. Hence, overall, Proposition \ref{prop2} implies that when either $n\leq m$ or $n>m$ together with $w_i\geq \frac{n-2}{m-2}$,  the use of WMIN in the scenario $x^{D,i}$ with the weights $(w_1,\ldots,w_n)\in W^{\mbox{\scriptsize WMIN}}$ corresponds to the use of WMEAN with the weights $(\beta_1,\ldots,\beta_n)\in W^{\mbox{\scriptsize WMEAN}}$ calculated according to Eq. \eqref{eq:wminOnwmean}. Under these circumstances, the results \eqref{eq:virhe1} and \eqref{eq:virhe2} concerning WMEAN can be also be utilized with WMIN.

\pagebreak\textbf{Proof of Proposition 2}

First, Eq. \eqref{eq:prop2aa} is proved.
Let $n\leq m$ and let a combination of points $(z_1^*,\ldots,z_n^*)$ be defined by 
\begin{equation}\left\{
\begin{array}{ll}
z_i^*=\frac{1}{m}\\
z_r^*=\frac{m-1}{m}\\
z_j^*=1~\forall~j=1,\ldots,n,~j\neq i,r
\end{array}\right..
\end{equation}
Then, it follows
\begin{equation}
\sum_{j=1}^n\frac{z_j^*}{n}=\frac{n-1}{n}\leq \frac{m-1}{m}.
\end{equation}
Let $(w_1,\ldots,w_n)\in W^{\mbox{\scriptsize WMIN}}$ (Eq. \eqref{eq:wmin}) and for $j=1,\ldots,n$, let the function $y_j(\cdot)$ be defined by
\begin{align}\label{eq:func_y}
&y_j(z_1,\ldots,z_n,w_j)=\frac{w_j z_j+\sum_{t\neq j}z_t}{w_j+n-1}=\frac{(w_j-1)z_j+n\sum_{t=1}^n\frac{z_t}{n}}{w_j-1+n}
\nonumber\\
\addlinespace
&=\alpha_j z_j+(1-\alpha_j)\sum_{t=1}^n\frac{z_t}{n},\quad \alpha_j=\frac{w_j-1}{w_j-1+n}.
\end{align}
Note that $w_j=1 \Leftrightarrow \alpha_j=0$ and $w_j\rightarrow \infty\Leftrightarrow \alpha_j\rightarrow 1$.
Thus, for all $n\geq 2$ and $m\geq 2$, it is obtained
\begin{align}\label{eq:noka}
&\left\{
\begin{array}{ll}
y_i(z_1^*,\ldots,z_n^*,w_i)=\alpha_i\frac{1}{m}+(1-\alpha_i)\frac{n-1}{n} \leq \frac{n-1}{n}\\
y_r(z_1^*,\ldots,z_n^*,w_r)=\alpha_r\frac{m-1}{m}+(1-\alpha_r)\frac{n-1}{n}\geq \frac{n-1}{n}\\
y_j(z_1^*,\ldots,z_n^*,w_j)=\alpha_j+(1-\alpha_j)\frac{n-1}{n}\geq \frac{n-1}{n}~\forall~j=1,\ldots,n,~j\neq i,r
\end{array}\right.\nonumber\\
&\Rightarrow y_i(z_1^*,\ldots,z_n^*,w_i)\leq y_j(z_1^*,\ldots,z_n^*,w_j)~\forall~j=1,\ldots,n,~j\neq i.
\end{align}
Consider then an arbitrary combination of points $(z_1,\ldots,z_n)$ such that
\begin{equation}\label{eq:arbZa}\left\{
\begin{array}{ll}
z_i\in[0,\frac{1}{m}]\\
z_j\in[\frac{m-1}{m},1]~\forall~j=1,\ldots,n,~j\neq i.
\end{array}\right.
\end{equation}
By using the notation 
\begin{equation}\label{eq:arbZb}\left\{
\begin{array}{ll}
z_i=z_i^*-e_i,~e_i\geq 0\\
z_r=z_r^*+e_r,~e_r\geq 0\\
z_j=z_j^*-e_j,~e_j\geq 0~\forall~j=1,\ldots,n,~j\neq i,r
\end{array}\right.,
\end{equation}
it can be derived
\begin{align}\label{eq:eka}
&\sum_{t=1}^n\frac{z_t}{n}-z_i=\sum_{t=1}^n\frac{z_t^*}{n}+\frac{1}{n}(-e_i+e_r-\sum_{j\neq i,r} e_j)-z_i^*+e_i \nonumber\\
&=\frac{n-1}{n}+\frac{1}{n}(-e_i+e_r-\sum_{j\neq i,r} e_j)-\frac{1}{m}+e_i\nonumber\\
&=\frac{1}{n}(n-1+(n-1)e_i+e_r-\sum_{j\neq i,r} e_j)-\frac{1}{m}\nonumber\\
&\geq \frac{1}{n}(n-1-(n-2)\frac{1}{m})-\frac{1}{m}\nonumber\\
&=\frac{(m-2)(n-1)}{mn}\geq 0\nonumber\\
&\Leftrightarrow z_i \leq \sum_{t=1}^n\frac{z_t}{n}\nonumber\\
&\Rightarrow y_i(z_1,\ldots,z_n,w_i)\leq \sum_{t=1}^n\frac{z_t}{n},
\end{align}
where the final implication follows from Eq. \eqref{eq:func_y}.

Furthermore, with any $k=1,\ldots,n$, $k\neq i,r$, it is derived
\begin{align}\label{eq:toka}
&z_k-\sum_{t=1}^n\frac{z_t}{n}=z_k^*-e_k-\sum_{t=1}^n\frac{z_t^*}{n}-\frac{1}{n}(-e_i+e_r-\sum_{j\neq i,r} e_j) \nonumber\\
&=1-e_k-\frac{n-1}{n}-\frac{1}{n}(-e_i+e_r-\sum_{j\neq i,r} e_j)\nonumber\\
&=\frac{1}{n}+\frac{1}{n}(e_i-e_r+\sum_{j\neq i,r,k} e_j-(n-1)e_k)\nonumber\\
&\geq \frac{1}{n}+\frac{1}{n}(0-\frac{1}{m}+0-(n-1)\frac{1}{m})\nonumber\\
&=\frac{1}{n}-\frac{1}{m}\geq 0\nonumber\\
&\Leftrightarrow z_k \geq \sum_{t=1}^n\frac{z_t}{n} \nonumber\\
&\Rightarrow y_k(z_1,\ldots,z_n,w_i)\geq \sum_{t=1}^n\frac{z_t}{n},
\end{align}
where the final implication follows from Eq. \eqref{eq:func_y}.

In addition, one can also derive
\begin{align}\label{eq:koka}
&z_r-\sum_{t=1}^n\frac{z_t}{n}=z_r^*+e_r-\sum_{t=1}^n\frac{z_t^*}{n}-\frac{1}{n}(-e_i+e_r-\sum_{j\neq i,r} e_j) \nonumber\\
&=\frac{m-1}{m}+e_r-\frac{n-1}{n}-\frac{1}{n}(-e_i+e_r-\sum_{j\neq i,r} e_j)\nonumber\\
&=\frac{m-1}{m}-\frac{n-1}{n}+\frac{1}{n}(e_i+(n-1)e_r+\sum_{j\neq i,r} e_j)\nonumber\\
&\geq \frac{m-1}{m}-\frac{n-1}{n}\geq 0 \nonumber\\
&\Leftrightarrow z_r \geq \sum_{t=1}^n\frac{z_t}{n}\nonumber\\
&\Rightarrow y_r(z_1,\ldots,z_n,w_i)\geq \sum_{t=1}^n\frac{z_t}{n},
\end{align}
where the final implication follows from Eq. \eqref{eq:func_y}.

Together, Eqs. \eqref{eq:func_y}, \eqref{eq:noka}, \eqref{eq:eka}, \eqref{eq:toka}, and \eqref{eq:koka} imply
\begin{align}
&n\leq m \Rightarrow y_i(z_1,\ldots,z_n,w_i)\leq y_j(z_1,\ldots,z_n,w_j)~\forall~j=1,\ldots,n,~j\neq i,\nonumber\\
& ~\forall~ (z_1,\ldots,z_n)~:~
\left\{
\begin{array}{ll}
z_i\in[0,\frac{1}{m}]\\
z_j\in[\frac{m-1}{m},1]~\forall~j=1,\ldots,n,~j\neq i 
\end{array}\right.,~\forall~(w_1,\ldots,w_n)\in W^{\mbox{\scriptsize WMIN}}\nonumber\\
\addlinespace
&\Leftrightarrow \min\limits_{j=1,\ldots,n}\left\{\frac{w_j \chi_j+\sum_{t\neq j}\chi_t}{w_j+n-1}\right\}\equiv \frac{w_i \chi_i+\sum_{t\neq i}\chi_t}{w_i+n-1}~\forall~(w_1,\ldots,w_n)\in W^{\mbox{\scriptsize WMIN}}\nonumber\\
&\Leftrightarrow \mu(w_1,\ldots,w_n)\equiv\mu_i(w_i)~\forall~(w_1,\ldots,w_n)\in W^{\mbox{\scriptsize WMIN}},
\end{align}
which concludes the proof of Eq. \eqref{eq:prop2aa}.

Next, Eq. \eqref{eq:prop2bb} is proved. Thus, let $n>m$ and first assume that $\mu(w_1,\ldots,w_n)\equiv\mu_i(w_i)~\forall~w_j\geq 1,~j=1\ldots,n,~j\neq i$. With $(z_1^*,\ldots,z_n^*)$ it is now obtained
\begin{equation}\label{eq:apu}
\sum_{j=1}^n\frac{z_j^*}{n}=\frac{n-1}{n}> \frac{m-1}{m}=z_r^*. 
\end{equation}
Based on the condition $\mu(w_1,\ldots,w_n)\equiv\mu_i(w_i)~\forall~w_j\geq 1,~j=1\ldots,n,~j\neq i$, it must apply with any $w_r\geq 1$ that $y_i(z_1^*,\ldots,z_n^*,w_i)\leq y_r(z_1^*,\ldots,z_n^*,w_r)$. With the aid of Eqs. \eqref{eq:func_y} and \eqref{eq:apu}, it is then obtained
\begin{align}\label{eq:apu2}
&y_i(z_1^*,\ldots,z_n^*,w_i)\leq y_r(z_1^*,\ldots,z_n^*,w_r)\nonumber\\
&\Leftrightarrow \alpha_i\frac{1}{m}-(1-\alpha_i)\frac{n-1}{n}\leq \alpha_r\frac{m-1}{m}-(1-\alpha_r)\frac{n-1}{n}= \frac{m-1}{m}+e,~e>0,~ e \xrightarrow{\alpha_r\rightarrow 1} 0\nonumber\\
&\Leftrightarrow\alpha_i \geq \frac{m-n}{m+n-mn}+\frac{mn*e}{m+n-mn}\nonumber\\
\addlinespace
&\Leftrightarrow w_i\geq \frac{n-2+m(1-n)*e}{m-2+m*e}\xrightarrow{\alpha_r\rightarrow 1} \frac{n-2}{m-2}.
 \end{align}
Thereby, it has been shown that
\begin{equation}
n>m ~\mbox{and}~\mu(w_1,\ldots,w_n)\equiv\mu_i(w_i)~\forall~w_j\geq 1,~j=1\ldots,n,~j\neq i \Rightarrow w_i\geq \frac{n-2}{m-2},
\end{equation} 
whereby half of the proof of Eq. \eqref{eq:prop2bb} is done.

To complete the proof of Eq. \eqref{eq:prop2bb}, let it now be so that $n>m$ and $w_i\geq \frac{n-2}{m-2}$. Furthermore, suppose there is a combination of points $(z_1,\ldots,z_n)$ and weights $(w_1,\ldots,w_n)\in W^{\mbox{\scriptsize WMIN}}$ such that
\begin{equation}
\left\{
\begin{array}{ll}
z_i\in[0,\frac{1}{m}]\\
z_j=\in[\frac{m-1}{m},1]~\forall~j=1,\ldots,n,~j\neq i\\
y_r(z_1,\ldots,z_n,w_r)< y_i(z_1,\ldots,z_n,w_i)
\end{array}\right..
\end{equation}
Referring to Eq. \eqref{eq:func_y}, note that now
\begin{align}\label{eq:painoapu}
&y_r(z_1,\ldots,z_n,w_r)< y_i(z_1,\ldots,z_n,w_i) \Leftrightarrow \alpha_r\underbrace{\left(\sum_{t=1}^n\frac{z_t}{n}-z_r\right)}_{=:d}>\alpha_i\underbrace{\left(\sum_{t=1}^n\frac{z_t}{n}-z_i\right)}_{\geq d}\nonumber\\
&\Rightarrow \alpha_r> \alpha_i \Leftrightarrow w_r > w_i.
\end{align}
Then, by writing the points $(z_1^*,\ldots,z_n^*)\neq (z_1,\ldots,z_n)$ as
\begin{equation}\label{eq:tildeChi}\left\{
\begin{array}{ll}
z_i^*=\frac{1}{m}=z_i+e_i,~e_i\geq 0 \\
z_r^*=\frac{m-1}{m}=z_r-e_r,~e_r\geq 0\\
z_j^*=1=z_j+e_j,~e_j\geq 0~\forall~j=1,\ldots,n,~j\neq i,r
\end{array}\right.,~\sum_{j=1}^n e_j>0,
\end{equation}
it can be derived based on Eq. \eqref{eq:painoapu} that
\begin{align}\label{eq:tildeChiero}
&y_i(z_1^*,\ldots,z_n^*,w_i)-y_r(z_1^*,\ldots,z_n^*,w_r)\nonumber\\
&=y_i(z_1,\ldots,z_n,w_i)-y_r(z_1,\ldots,z_n,w_r)\nonumber\\
&+e_i\underbrace{\left(\frac{w_i}{w_i+n-1}-\frac{1}{w_r+n-1}\right)}_{> 0}+e_r\underbrace{\left(\frac{w_r}{w_r+n-1}-\frac{1}{w_i+n-1}\right)}_{> 0}\nonumber\\
&+\sum_{j\neq i,r}e_j\underbrace{\left(\frac{1}{w_i+n-1}-\frac{1}{w_r+n-1}\right)}_{>0}\nonumber\\
&> 0 \Leftrightarrow y_i(z_1^*,\ldots,z_n^*,w_i)>y_r(z_1^*,\ldots,z_n^*,w_r).
\end{align}

However, given the condition $w_i \geq \frac{n-2}{m-2}$, Eq. \eqref{eq:apu2} implies that $y_i(z_1^*,\ldots,z_n^*,w_i)\leq y_r(z_1^*,\ldots,z_n^*,w_r)$ which contradicts Eq. \eqref{eq:tildeChiero}. Thereby, referring to Eqs. \eqref{eq:func_y} and \eqref{eq:muexpres}, it is concluded that 
\begin{align}
&n>m~\mbox{and}~w_i\geq \frac{n-2}{m-2}\nonumber\\
&\Rightarrow \nonumber\\
&y_i(z_1,\ldots,z_n,w_i)\leq y_j(z_1,\ldots,z_n,w_j)~\forall~w_j\geq 1,~j=1,\ldots,n,~j\neq i, \nonumber\\
&\forall~ (z_1,\ldots,z_n)~:~
\left\{
\begin{array}{ll}
z_i\in[0,\frac{1}{m}]\\
z_j\in[\frac{m-1}{m},1]~\forall~j=1,\ldots,n,~j\neq i
\end{array}\right.\nonumber\\
\addlinespace
&\Leftrightarrow \min\limits_{j=1,\ldots,n}\left\{\frac{w_j \chi_j+\sum_{t\neq j}\chi_t}{w_j+n-1}\right\}\equiv \frac{w_i \chi_i+\sum_{t\neq i}\chi_t}{w_i+n-1}\nonumber\\
&\Leftrightarrow \mu(w_1,\ldots,w_n)\equiv\mu_i(w_i)~\forall~w_j\geq 1,~j=1,\ldots,n,~j\neq i,
\end{align}
which concludes the proof of Eq. \eqref{eq:prop2bb}.

\section{Experiment on weight intervals for MIXMINMAX}\label{sec:MIXstudy}

In Section \ref{sec:WMEANProps}, it was shown that when a probability distribution $P(X_C|x^{D,i})$ is generated with WMEAN, a single weight $w_i$ basically solely defines the mode pair $p^{D,i}$ of the child node $X_C$. Furthermore, through Eqs. \eqref{eq:virhe1} and \eqref{eq:virhe2}, one can determine for $w_i$ bounds $\underline{w}_i^D$ and $\overline{w}_i^D$ so that the interval $[\underline{w}_i^D, \overline{w}_i^D]$  represents a good approximation of all the values of $w_i$ by some consecutive states $x_C^a$ and $x_C^b$ of  $X_C$ form $p^{D,i}$. When $P(X_C|x^{D,i})$  is generated with MIXMINMAX, see Eq. \eqref{eq:mixminmax}, the distribution can also be seen to depend only on a single weight. This can be either $w_{MIN}$ or $w_{MAX}$ as the two weights are bound together by $w_{MIN}+w_{MAX}=1$.  Thus, if the single weight is selected to be $w_{MAX}$ , the use of $w_{MIN}$ can be linked to $w_{MAX}$ with the substitution $w_{MIN}=1-w_{MAX}$. 

Concerning the mode pair $p^{D,i}$ of $P(X_C|x^{D,i})$ generated with MIXMINMAX, one can determine an interval $[\underline{w}_{MAX}^{D,i}, \overline{w}_{MAX}^{D,i}]$ such that when $w_{MAX}\in [\underline{w}_{MAX}^{D,i}, \overline{w}_{MAX}^{D,i}]$, $P(X_C|x^{D,i})$ is compatible with  $p^{D,i}$. The interval can be determined, e.g., with the bisection method \cite{bazaraa2006nonlinear}. In this section, it experimentally shown that the variance parameter $\sigma^2$ and the sample size parameter $s$ used in the bisection method have small effect to the bounds $\underline{w}_{MAX}^{D,i}$ and $\overline{w}_{MAX}^{D,i}$. 

\subsection{Experimental setup}

The effects of $\sigma^2$ and $s$ to the values of $\underline{w}_{MAX}^{D,i}$ and $\overline{w}_{MAX}^{D,i}$ are studied through the functions $D_{1}$ and $D_{2}$ defined as
\begin{align}
&D_{1}(n,m,\boldsymbol{w},\sigma^2,s,k)=\left|P(X_C=x_C^k|x^{D,i}; \MIXMINMAX,\boldsymbol{w},\sigma^2,s)\right.\nonumber\\
&\quad\qquad\qquad\qquad\qquad\qquad\left.-P(X_C=x_C^{k+1}|x^{D,i}; \MIXMINMAX,\boldsymbol{w},\sigma^2,s)\right|,\nonumber\\
&D_{2}(n,m,\boldsymbol{w},\sigma^2,s,k)=\left|P(X_C=x_C^{k-1}|x^{D,i}; \MIXMINMAX,\boldsymbol{w},\sigma^2,s)\right.\nonumber\\
&\quad\qquad\qquad\qquad\qquad\qquad\left.-P(X_C=x_C^{k+1}|x^{D,i}; \MIXMINMAX,\boldsymbol{w},\sigma^2,s)\right|.
\end{align} 
Now, $D_{1}$ is the difference of the probabilities of consecutive states $x_C^k$ and $x_C^{k+1}$ when the probability distribution $P(X_C|x^{D,i})$ is generated in a setting of $n$ parent nodes and $m$ states of the nodes using MIXMINMAX with the weights $\boldsymbol{w}=(1-w_{MAX},w_{MAX})$,  the variance parameter $\sigma^2$, and the sample size $s$. The function $D_{2}$ portrays the difference between the probabilities of states $x_C^{k-1}$ and $x_C^{k+1}$ in the same setting. Note that with both $D_{1}$ and $D_{2}$, $n$ (number of parent nodes) and $m$ (number of states of the nodes) are among the arguments even though their presence is not explicitly shown in the expressions defining the functions. However, their presence in the functions is evident based on the functioning of RNM explained in Section \ref{sec:CPDGen}. 

The experiment proceeds as follows. First, for fixed values of $n$, $m$, $s$, and $k$ as well as for a selected value of the variance parameter $\sigma^2_0$, the bisection method is used to determine a specific value of the weight $w_{MAX}$, denoted by $w_{MAX}^0$. The weight $w_{MAX}^0$ is such that with $\boldsymbol{w}^0=(1-w_{MAX}^0,w_{MAX}^0)$, $D_{j}(n,m,\boldsymbol{w}^0,\sigma^2_0,s,k)=0$ ($j=1$ or $j=2$).  After $w_{MAX}^0$ is determined,  function values $D_{j}(n,m,\boldsymbol{w}^0,\sigma^2,s,k)$ are calculated for multiple different values of $\sigma^2$. Small values of $D_{j}(n,m,\boldsymbol{w}^0,\sigma^2,s,k)$ indicate that $\sigma^2$ has small effect to the bounds of the elicitation weight intervals. The experiment includes repetitions in which different values of $n$, $m$, $s$, and $k$ are used.

\subsection{Results}

\begin{figure*}[!h]
\centering
\subfloat[]{\includegraphics[scale=0.4]{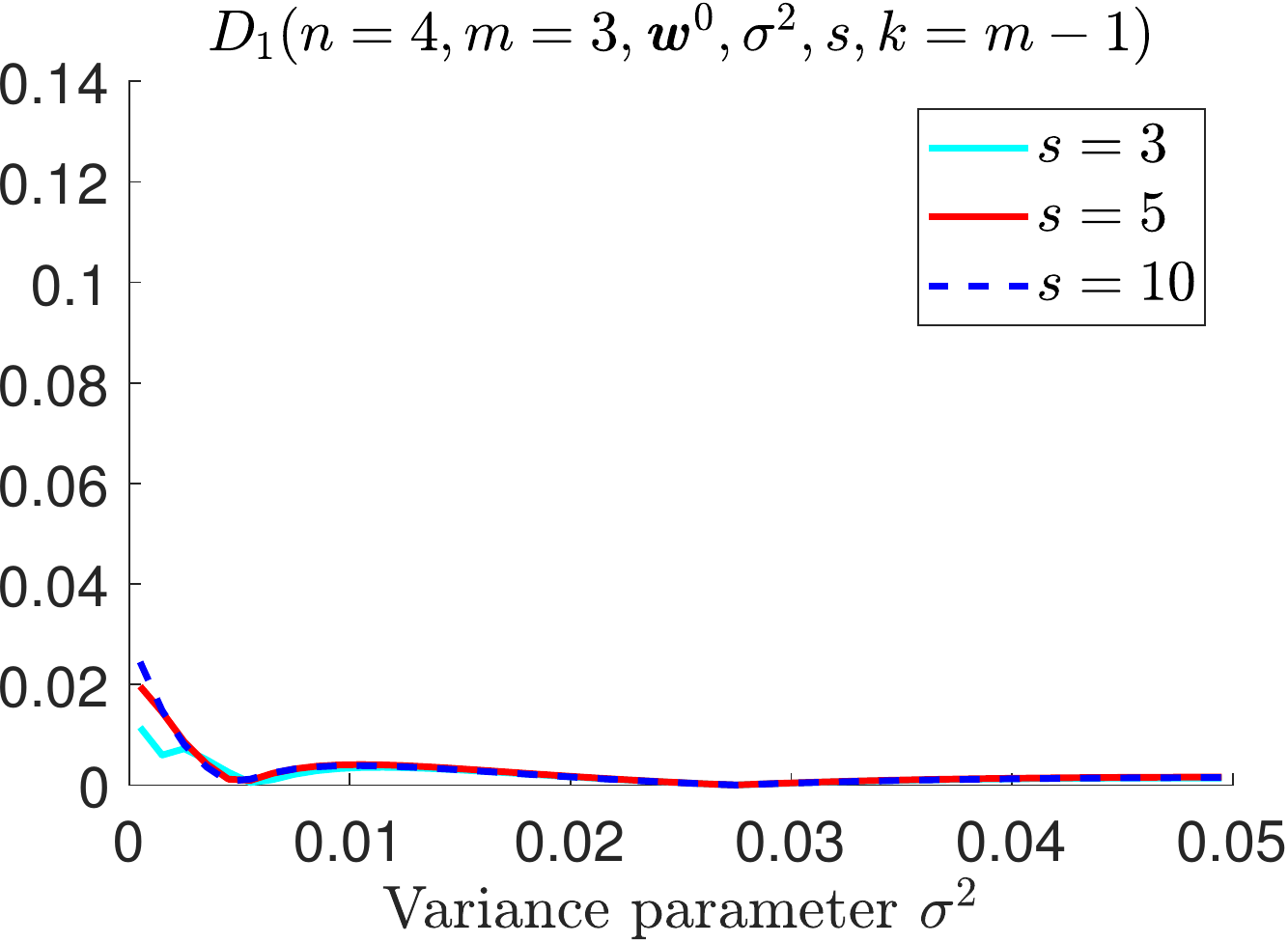}}
~~~
\subfloat[]{\includegraphics[scale=0.4]{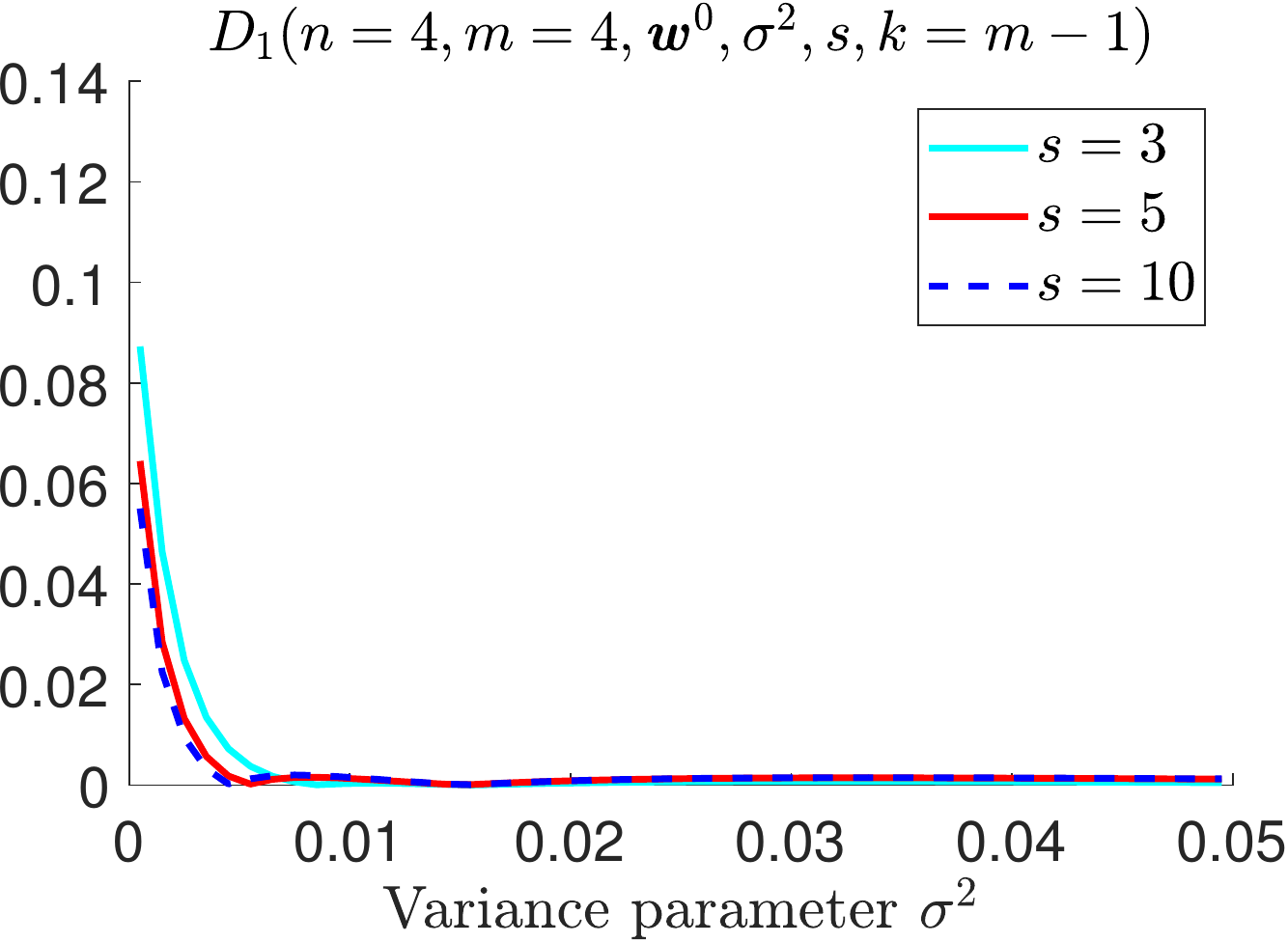}}
~~~
\subfloat[]{\includegraphics[scale=0.4]{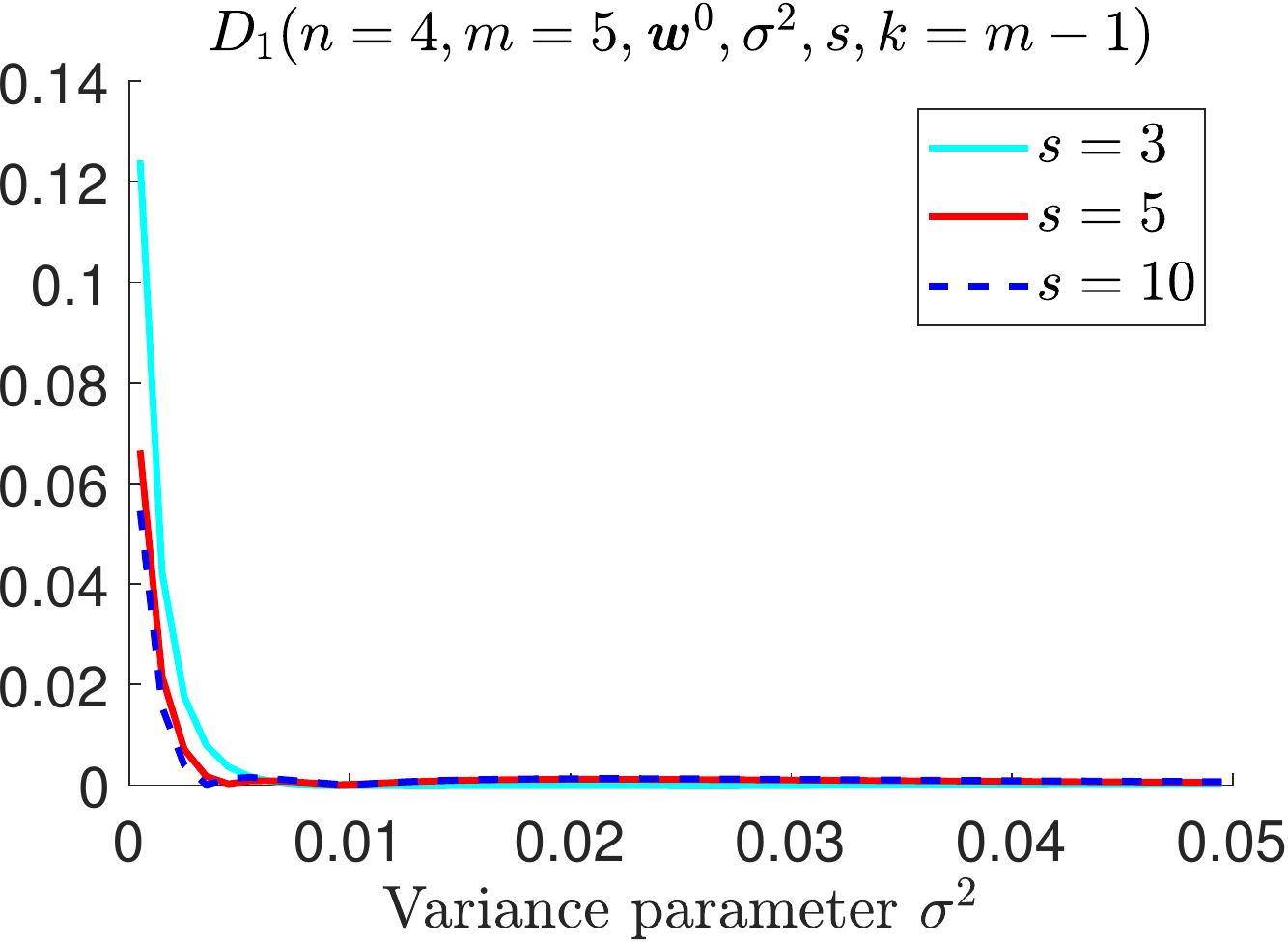}}
\vfill
\subfloat[]{\includegraphics[scale=0.4]{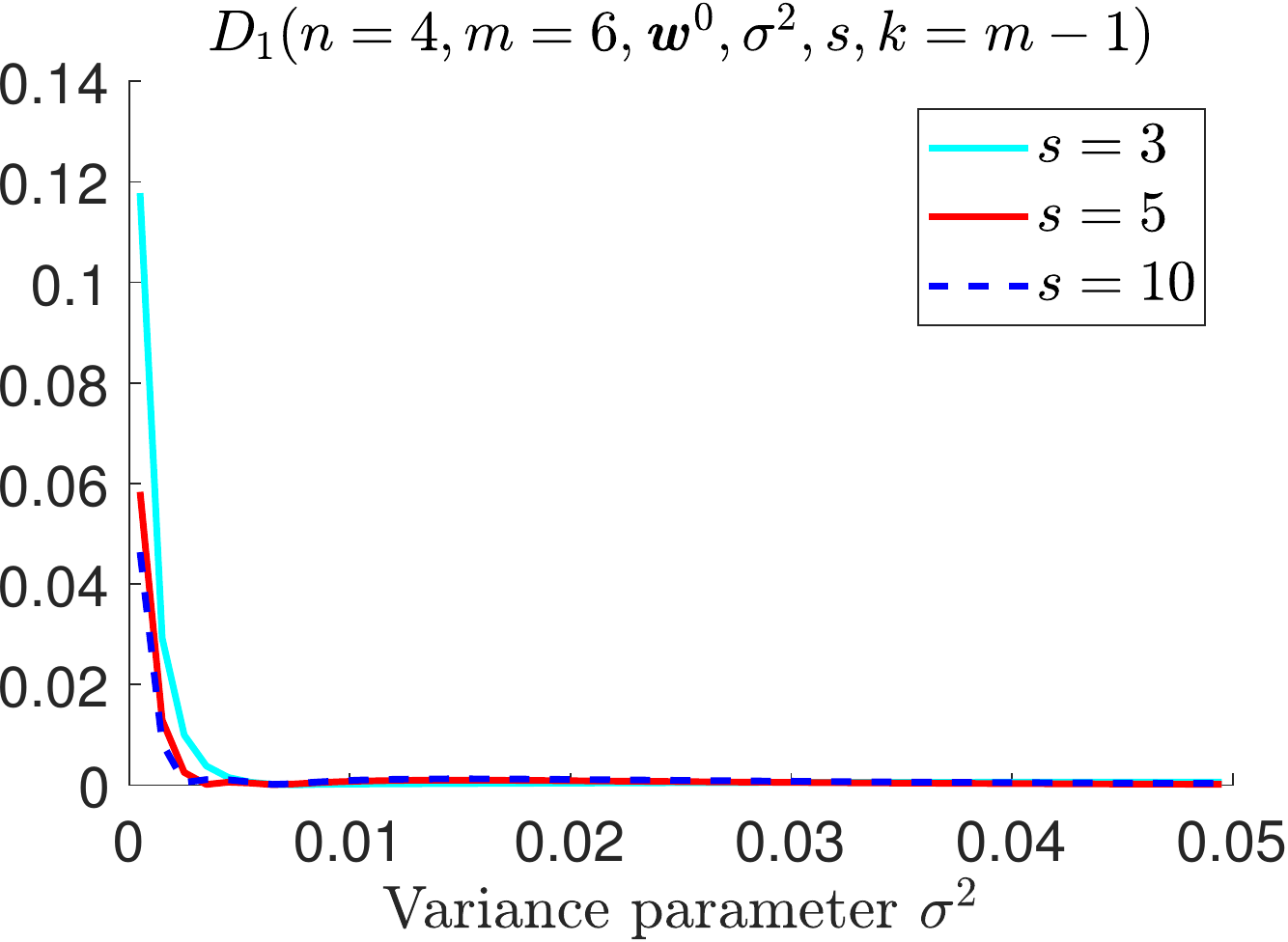}}
~~~
\subfloat[]{\includegraphics[scale=0.4]{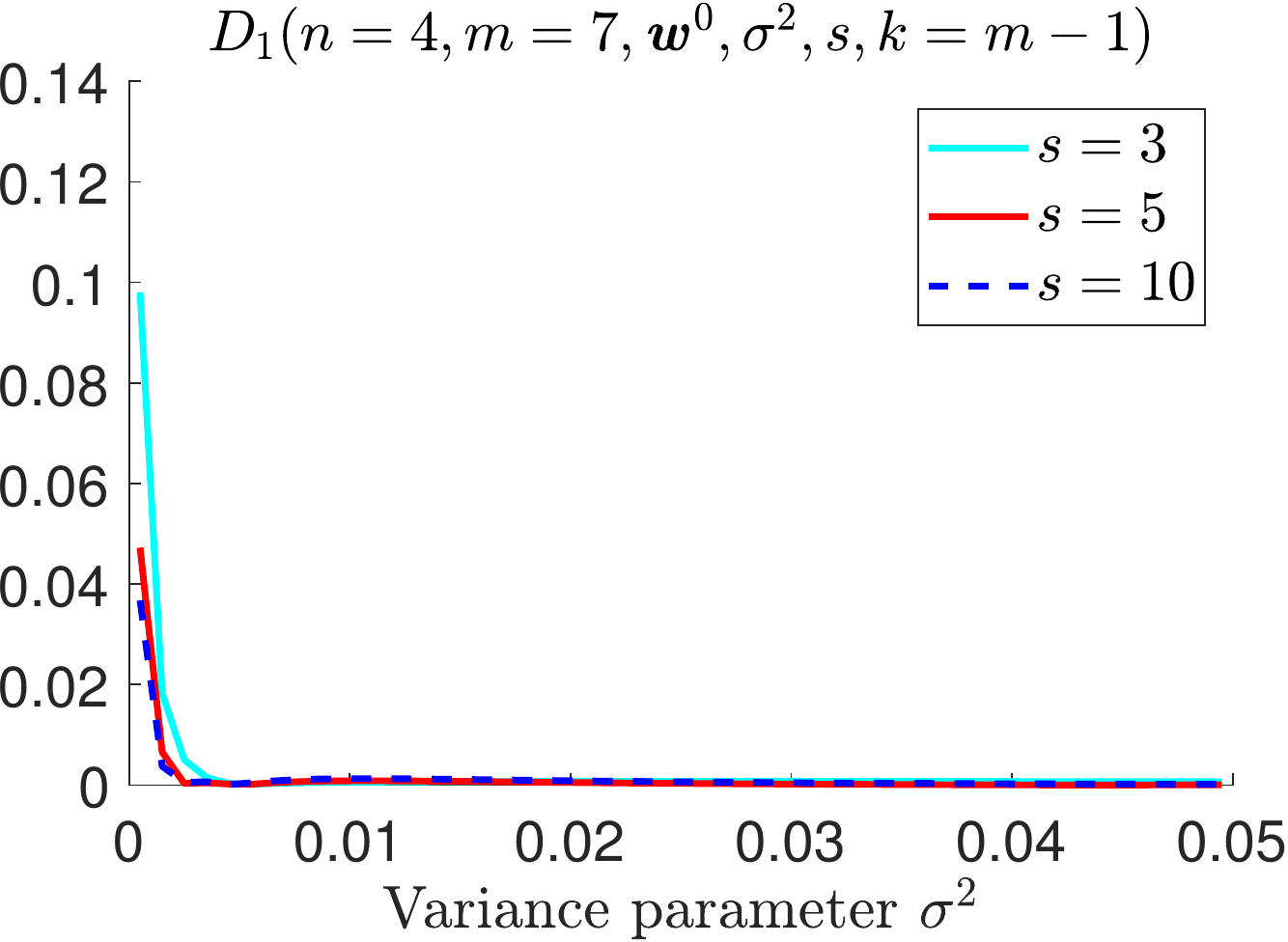}}
~~~
\subfloat[]{\includegraphics[scale=0.4]{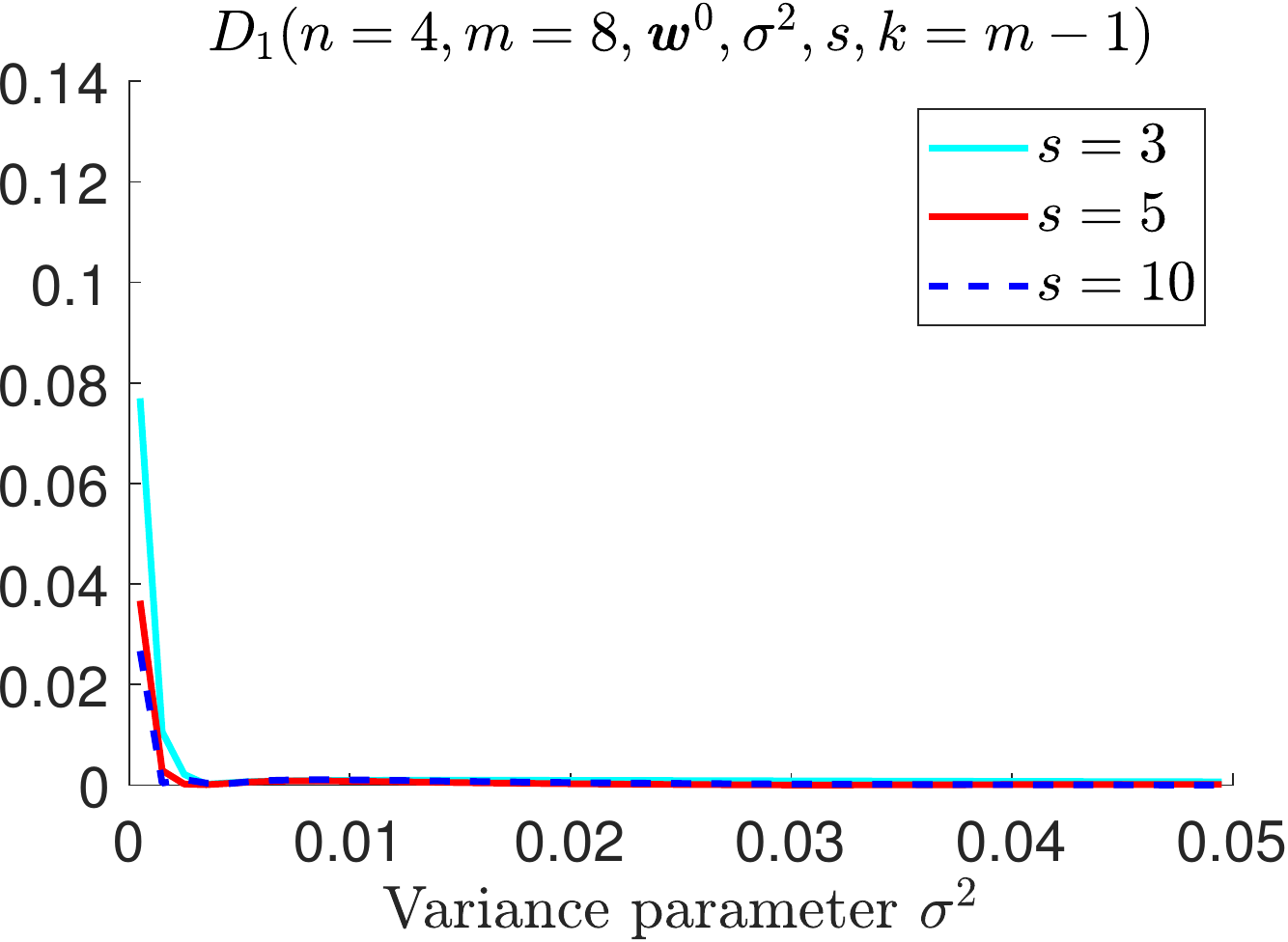}}
\vfill
\subfloat[]{\includegraphics[scale=0.4]{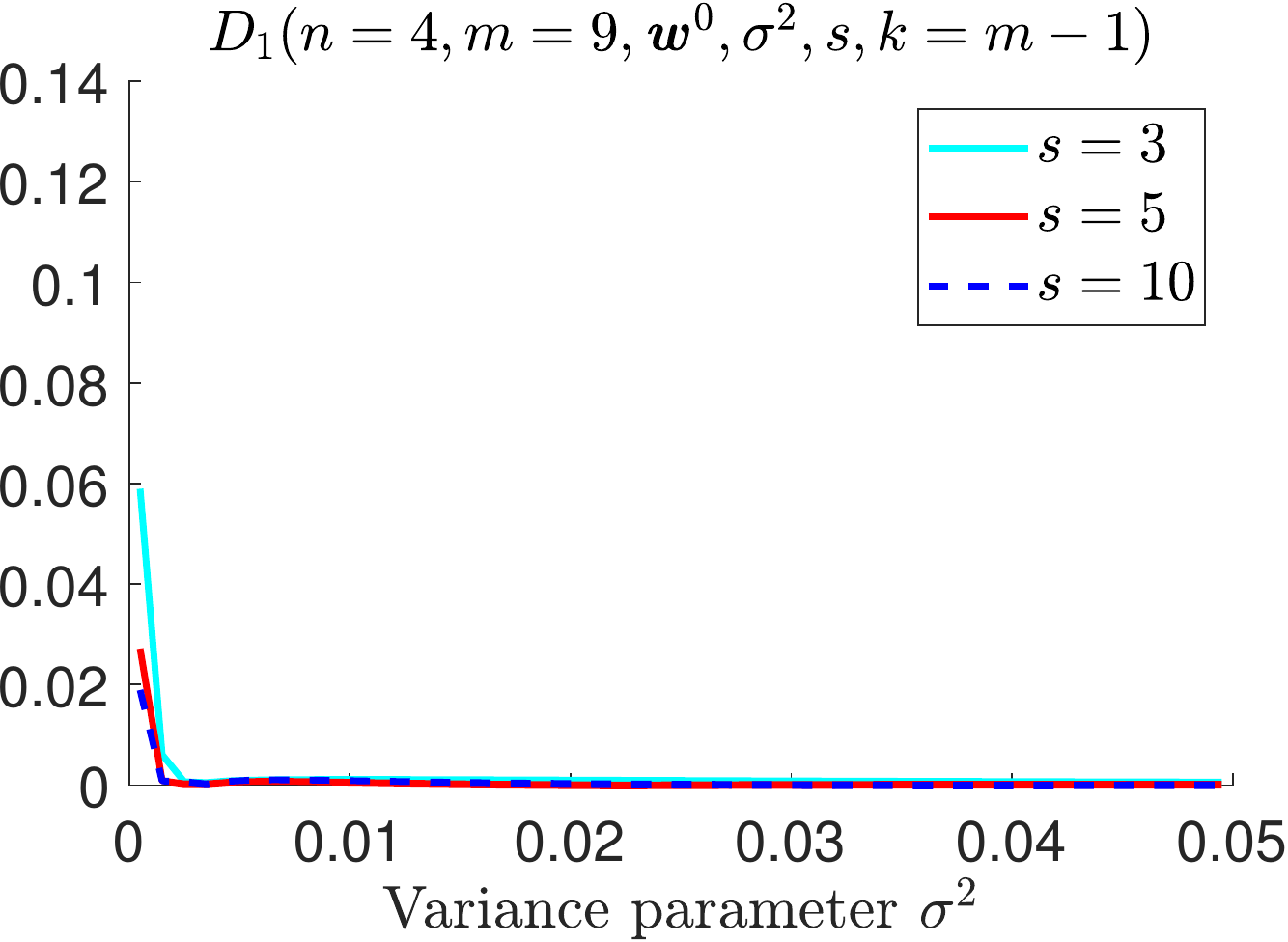}}
~~~
\subfloat[]{\includegraphics[scale=0.4]{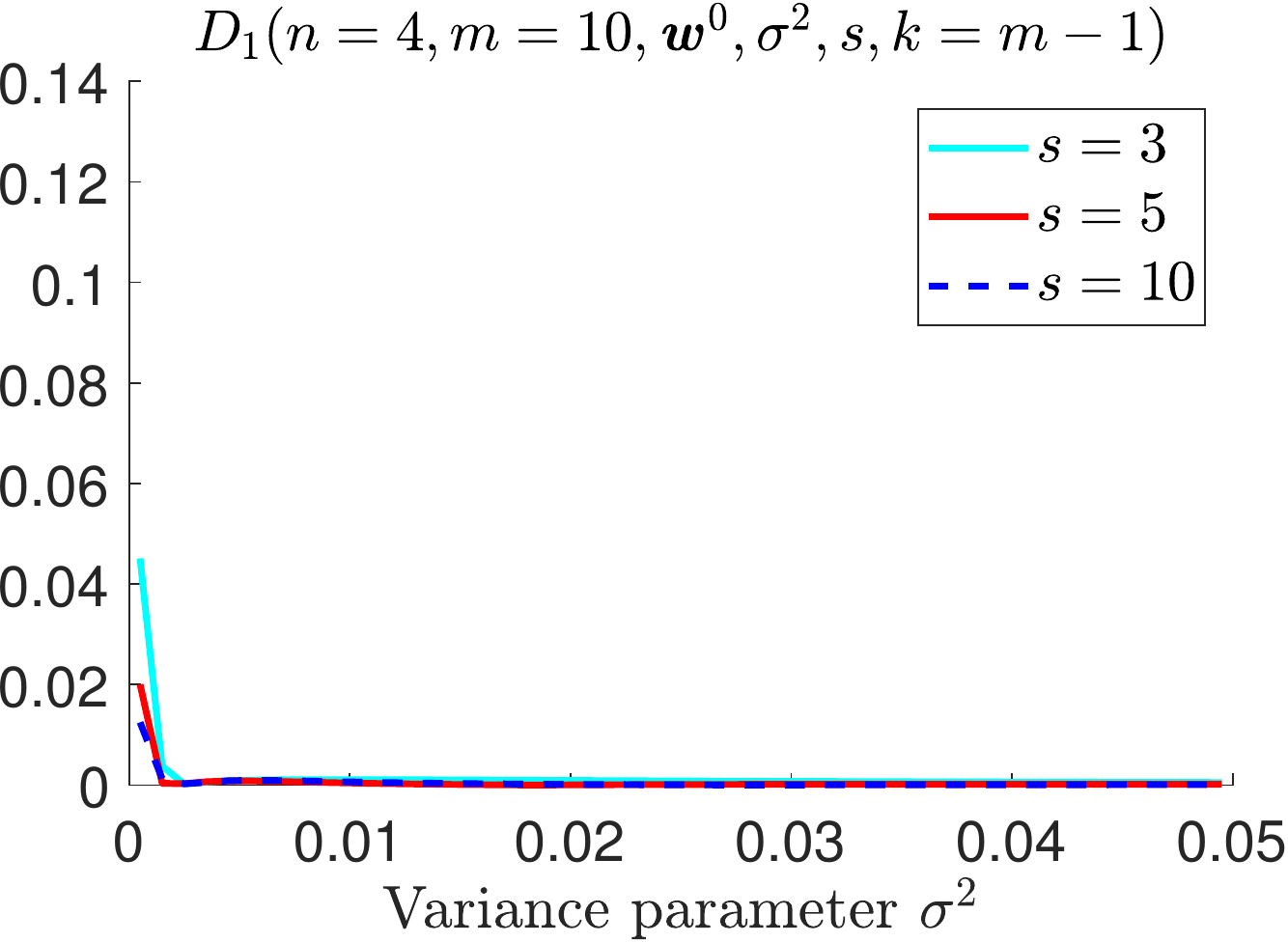}}
~~~
\subfloat[]{\includegraphics[scale=0.4]{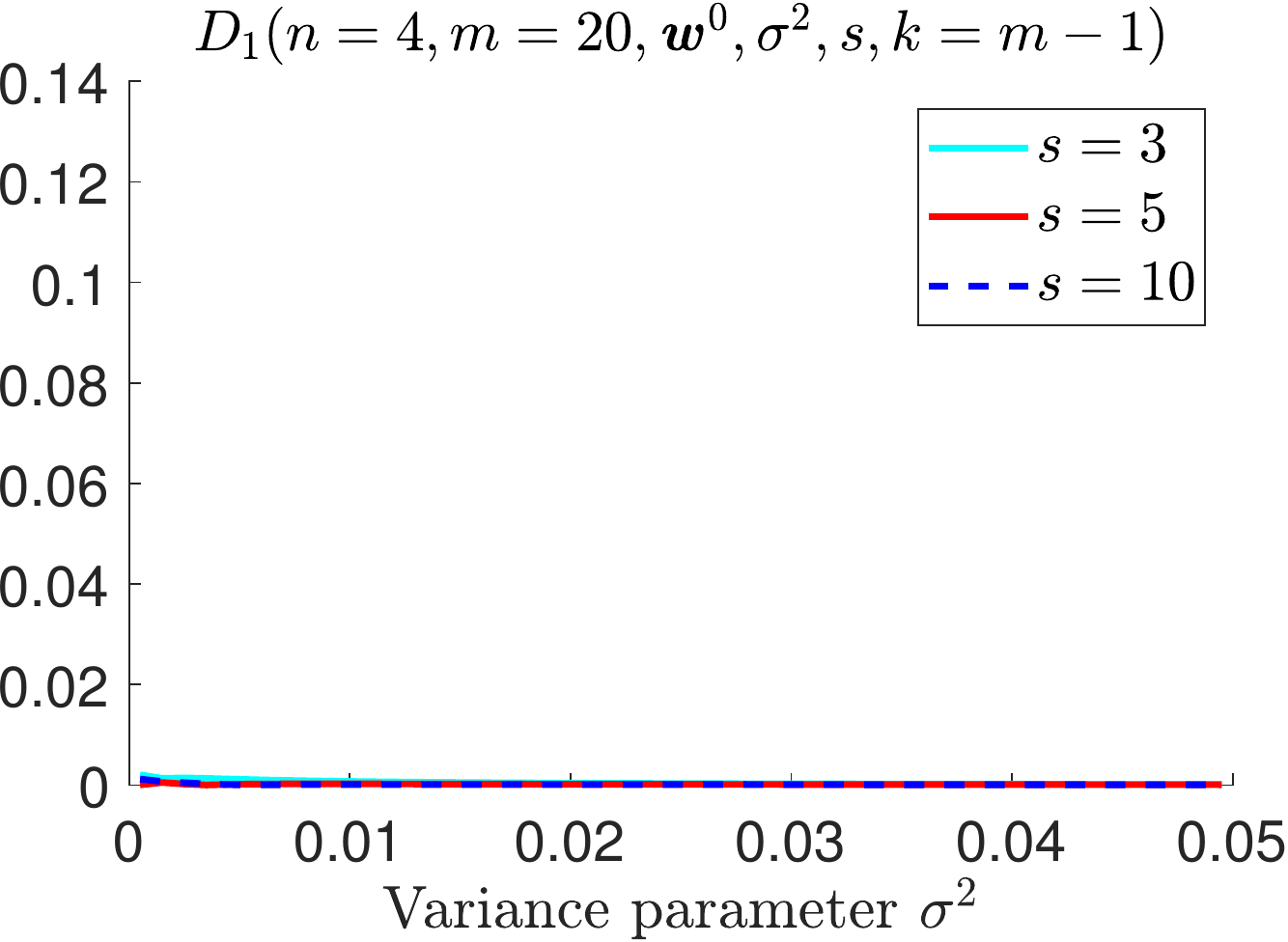}}
\caption{The function $D_{1}(n=4,m,\boldsymbol{w}^0,\sigma^2,s,k=m-1)$ plotted as a function of the variance parameter $\sigma^2$ with $s=3$, $s=5$, and $s=10$, and (a) $m=3$, (b) $m=4$, (c) $m=5$, (d) $m=6$, (e) $m=7$, (f) $m=8$, (g) $m=9$, (h) $m=10$, and (i) $m=20$.}
\label{fig:MIXapp}
\end{figure*}

When the functions $D_{1}$ and $D_{2}$ are tentatively numerically explored with different values of $n$, $m$, and $k$ (the results are not presented here), it is discovered that $D_{1}$ provides the largest values. Concerning $D_{1}$, the largest values are always obtained when $k=m-1$. Moreover, with regard to different values of $n\geq 2$, the largest single values of $D_{1}$ are obtained in the case $n=4$. 

Based on the results of the tentative exploration, the focus is now set only on the function $D_{1}$ and the cases $n=4$ and $k=m-1$ in order to get idea of the maximum effect that $\sigma^2$ and $s$ can have on the values of $\underline{w}_{MAX}^{D,i}$ and $\overline{w}_{MAX}^{D,i}$ determined with the bisection method. Fig. \ref{fig:MIXapp} displays $D_{1}(n=4,m,\boldsymbol{w}^*,\sigma^2,s,k=m-1)$ as a function of $\sigma^2$ for various values of $m$ when $\sigma^2_0=1/(4m^2)$ and $s=3,5,10$. With $\sigma^2=1/(4m^2)$, the majority of the probability mass in a probability distribution $P(X_C|x^{D,i}; \MIXMINMAX,\boldsymbol{w},\sigma^2,s)$ is shared between the two or three most probable states of the child node, cf. the discussion concerning Eq. \eqref{eq:int095}. The values $s=3$ and $s=5$ are the smallest allowed and the default values of $s$ in AgenaRisk software \cite{agenarisk10}, respectively.

In each of the graphs displayed in Fig. \ref{fig:MIXapp}, $D_{1}<0.01$ with most values of $\sigma^2$. Only when $\sigma^2$ approaches zero, the values of $D_{1}$ rapidly increase. However, this does not mean that changing the value of $\sigma^2_0$ from $\sigma^2_0=1/(4m^2)$ to a smaller value would lead to a large absolute change in the value of $w_{MAX}^0$. Rather, it reflects the fact that with a small value of $\sigma^2$, the probabilities generated with RNM generally become more sensitive to changes in the value of the weight $w_{MAX}$. For example, with $m=5$, $\sigma^2_0=1/(4m^2)=1/(4*5^2)=0.01$, and $s=3$, the value $w_{MAX}^0$ determined with the bisection method is $w_{MAX}^0=0.197$. Then, with $\sigma^2=5*10^{-4}$ and $\boldsymbol{w}=(1-0.197,0.197)$, the function $D_{R,1}$ obtains a value $D_{R,1}=0.124$ which one could interpret to indicate that the value of $w_{MAX}^0$ determined with the bisection method is highly sensitive to the value of $\sigma^2_0$ used. Yet, with $\sigma^2_0=5*10^{-4}$, the value of $w_{MAX}^0$ determined with the bisection method becomes $w_{MAX}^0=0.205$ which is close to the original value $w_{MAX}^0=0.197$.

In all of the graphs displayed in Fig. \ref{fig:MIXapp}, the curves corresponding to cases $s=5$ and $s=10$ are almost identical. The curve corresponding to $s=3$ also always tends to be similar to the other two except for the smallest values of $\sigma^2$. When $\sigma^2$ gets small enough, the $D_{1}$ values obtained with $s=3$ become clearly larger than those obtained with $s=5$ and $s=10$. However, this does not mean that the values of $w_{MAX}^0$ obtained with different sample sizes would differ from each other a lot. For example, with $m=5$ and $\sigma^2_0=5*10^{-4}$, the values of $w_{MAX}^0$ solved with the bisection method are $w_{MAX}^0=0.205$, $w_{MAX}^0=0.193$, and $w_{MAX}^0=0.187$, for $s=3$, $s=5$, and $s=10$, respectively. As discussed above, the phenomenon of $D_{1}$ values rising with small values of $\sigma^2$ is basically the result of the fact that the probabilities generated with RNM generally become more sensitive to changes in the value of $w_{MAX}$ when $\sigma^2$ is small. In the case of $s=3$, this phenomenon is just stronger than with $s=5$ and $s=10$. 

Based on the results of the experiment, it is concluded that when an interval $[\underline{w}_{MAX}^{D,i},\overline{w}_{MAX}^{D,i}]$ is determined for the weight $w_{MAX}$ with the bisection method with a fixed value of the variance parameter $\sigma^2$, the actual value of $\sigma^2$ does not affect the bounds $\underline{w}_{MAX}^{D,i}$ and $\overline{w}_{MAX}^{D,i}$ a lot from the practical point of view. That is, the values of  $\underline{w}_{MAX}^{D,i}$ and $\overline{w}_{MAX}^{D,i}$ determined with one value of $\sigma^2$ approximate well the critical values of $w_{MAX}$ by which $P(X_C|x^{D,i})$ stays compatible with a given mode pair $p^{D,i}$, no matter what value of $\sigma^2$ is used to generate $P(X_C|x^{D,i})$. Furthermore, the robustness of $\underline{w}_{MAX}^{D,i}$ and $\overline{w}_{MAX}^{D,i}$  for different values of $\sigma^2$ benefits from the sample size parameter of RNM being large enough, e.g., $s=5$. 

\section{Conclusion}\label{sec:conclusion}
This paper established on RNM technical results through means of analytical and experimental studying. Majority of the results are focused on a setting in which a child node and its parent nodes all have the same amount of states. The results can be used to support any further elaboration and development of RNM. For example, the results can help to develop new means to support elicitation of RNM parameters from a domain expert in practical applications. In future research, one could try to generalize the results presented in this paper for a setting in which the nodes have varying numbers of states.


\begin{thebibliography}{10}
\expandafter\ifx\csname url\endcsname\relax
  \def\url#1{\texttt{#1}}\fi
\expandafter\ifx\csname urlprefix\endcsname\relax\def\urlprefix{URL }\fi
\expandafter\ifx\csname href\endcsname\relax
  \def\href#1#2{#2} \def\path#1{#1}\fi

\bibitem{fenton2013risk}
N.~Fenton, M.~Neil, {Risk Assessment and Decision Analysis with Bayesian
  Networks}, CRC Press, Boca Raton, FL, USA, 2013.

\bibitem{russell2003artificial}
S.~Russell, P.~Norvig, {Artificial Intelligence: A Modern Approach}, Prentice
  Hall, Upper Saddle River, NJ, USA, 2003.

\bibitem{neapolitan2004learning}
R.~Neapolitan, {Learning Bayesian Networks}, Pearson Prentice Hall, Upper
  Saddle River, NJ, USA, 2004.

\bibitem{hill2021bayesian}
A.~Hill, C.~H. Joyner, C.~Keith-Jopp, B.~Yet, C.~T. Sakar, W.~Marsh,
  D.~Morrissey, A {B}ayesian network decision support tool for low back pain
  using a {RAND} appropriateness procedure: Proposal and internal pilot study,
  JMIR Research Protocols 10~(1) (2021) e21804/1--11.

\bibitem{constantinou2015causal}
A.~C. Constantinou, M.~Freestone, W.~Marsh, J.~Coid, Causal inference for
  violence risk management and decision support in forensic psychiatry,
  Decision Support Systems 80 (2015) 42--55.

\bibitem{heckerman1992toward}
D.~Heckerman, E.~Horvitz, B.~Nathwani, Toward normative expert systems: {P}art
  i. {T}he pathfinder project, Methods of Information in Medicine 31~(2) (1992)
  90--105.

\bibitem{topuz2021probabilistic}
K.~Topuz, D.~Delen, A probabilistic {B}ayesian inference model to investigate
  injury severity in automobile crashes, Decision Support Systems (2021)
  e113557/1--13.

\bibitem{tosoni2019comprehensiveness}
E.~Tosoni, A.~Salo, J.~Govaerts, E.~Zio, Comprehensiveness of scenarios in the
  safety assessment of nuclear waste repositories, Reliability Engineering \&
  System Safety 188 (2019) 561--573.

\bibitem{mancuso2021optimal}
A.~Mancuso, M.~Compare, A.~Salo, E.~Zio, Optimal {P}rognostics and {H}ealth
  {M}anagement-driven inspection and maintenance strategies for industrial
  systems, Reliability Engineering \& System Safety 210 (2021) e107536/1--10.

\bibitem{barons2017eliciting}
M.~J. Barons, S.~K. Wright, J.~Q. Smith, Eliciting probabilistic judgements for
  integrating decision support systems, in: L.~C. Dias, A.~Morton, J.~Quigley
  (Eds.), Elicitation: The Science and Art of Structuring Judgement, Springer,
  New York, 2017.

\bibitem{druzdel2000building}
M.~Druzdel, L.~van~der Gaag, Building probabilistic networks:"{W}here do the
  numbers come from?", IEEE Transactions on Knowledge and Data Engineering
  12~(4) (2000) 481--486.

\bibitem{monti2000dealing}
S.~Monti, G.~Carenini, Dealing with the expert inconsistency in probability
  elicitation, IEEE Transactions on Knowledge and Data Engineering 12~(4)
  (2000) 499--508.

\bibitem{rohmer2020uncertainties}
J.~Rohmer, Uncertainties in conditional probability tables of discrete
  {B}ayesian belief networks: A comprehensive review, Engineering Applications
  of Artificial Intelligence 88 (2020) 103384/1--16.

\bibitem{fenton2007using}
N.~Fenton, M.~Neil, J.~Caballero, Using ranked nodes to model qualitative
  judgments in {B}ayesian networks, IEEE Transactions on Knowledge and Data
  Engineering 19~(10) (2007) 1420--1432.

\bibitem{agenarisk10}
{Agena Ltd}, {AgenaRisk Software, Version 10.0}, \url{http://www.agenarisk.com}
  (February 2021).

\bibitem{kaya2019building}
R.~Kaya, B.~Yet, Building {B}ayesian networks based on {DEMATEL} for multiple
  criteria decision problems: A supplier selection case study, Expert Systems
  with Applications 134 (2019) 234--248.

\bibitem{xia2018hybrid}
N.~Xia, P.~X. Zou, X.~Liu, X.~Wang, R.~Zhu, A hybrid {BN-HFACS} model for
  predicting safety performance in construction projects, Safety science 101
  (2018) 332--343.

\bibitem{freire2018bayesian}
A.~Freire, M.~Perkusich, R.~Saraiva, H.~Almeida, A.~Perkusich, A {B}ayesian
  networks-based approach to assess and improve the teamwork quality of agile
  teams, Information and Software Technology 100 (2018) 119--132.

\bibitem{yet2016bayesian}
B.~Yet, A.~Constantinou, N.~Fenton, M.~Neil, E.~Luedeling, K.~Shepherd, A
  {B}ayesian network framework for project cost, benefit and risk analysis with
  an agricultural development case study, Expert Systems with Applications 60
  (2016) 141--155.

\bibitem{laitila2016improving}
P.~Laitila, K.~Virtanen, Improving construction of conditional probability
  tables for ranked nodes in {B}ayesian networks, IEEE Transactions on
  Knowledge and Data Engineering 28~(7) (2016) 1691--1705.

\bibitem{manu4}
P.~Laitila, K.~Virtanen, Portraying probabilistic relationships of continuous
  nodes in {B}ayesian networks with ranked nodes method, submitted for
  publication.

\bibitem{laitila2018ontheore}
P.~Laitila, K.~Virtanen, On theoretical principle and practical applicability
  of ranked nodes method for constructing conditional probability tables of
  {B}ayesian networks, IEEE Transactions on Systems, Man, and Cybernetics:
  Systems 50~(5) (2020) 1943--1955.

\bibitem{noguchi2019addressing}
T.~Noguchi, N.~Fenton, M.~Neil, Addressing the practical limitations of
  noisy-{OR} using conditional inter-causal anti-correlation with ranked nodes,
  IEEE Transactions on Knowledge and Data Engineering 31~(4) (2019) 813--817.

\bibitem{knapp2005basic}
A.~Knapp, {Basic Real Analysis}, Vol.~10, Birkh{\"a}user, Boston, MA, USA,
  2005.

\bibitem{kamgar-parsi1995distribution}
B.~Kamgar-Parsi, B.~Kamgar-Parsi, M.~Brosh, Distribution and moments of the
  weighted sum of uniform random variables, with applications in reducing
  {M}onte {C}arlo simulations, Journal of Statistical Computation and
  Simulation 52~(4) (1995) 399--414.

\bibitem{bazaraa2006nonlinear}
M.~Bazaraa, H.~Sherali, C.~Shetty, Nonlinear Programming: Theory and
  Algorithms, 3rd edition, John Wiley and Sons, Hoboken, NJ, USA, 2006.

\end{thebibliography}

\end{document}